\documentclass[12pt]{article}

\usepackage[margin=1.0in]{geometry}

\usepackage{cite}
\usepackage{booktabs}
\usepackage{amssymb,latexsym}

\usepackage{graphics,epsfig}

\usepackage{color}
\usepackage{xcolor}

\usepackage{amsmath,amsfonts}

\usepackage{mathtools}

\usepackage{mathrsfs}

\usepackage{slashed}

\usepackage{physics}
\usepackage{tensor}
\usepackage{graphicx}
\usepackage{float}
\usepackage{xcolor, colortbl}
\usepackage{hyperref}

\usepackage{authblk}

\DeclareMathAlphabet{\mathpzc}{OT1}{pzc}{m}{it}

\usepackage{listings}
\usepackage{color}

\definecolor{dkgreen}{rgb}{0,0.6,0}
\definecolor{gray}{rgb}{0.5,0.5,0.5}
\definecolor{mauve}{rgb}{0.58,0,0.82}

\usepackage{feynmp}
\usepackage{feynmp-auto}

\let\a=\alpha \let\b=\beta \let\g=\gamma \let\d=\delta \let\e=\epsilon
\let\z=\zeta  \let\th=\theta  \let\k=\kappa
\let\l=\lambda \let\m=\mu \let\n=\nu \let\x=\xi \let\p=\pi 
\let\s=\sigma   \let\f=\phi  
 
        \let\Th=\Theta \let\L=\Lambda
\let\X=\Xi  \let\S=\Sigma  \let\Y=\Psi
 
\let\la=\label  
  
\def\nn{\nonumber} \def\bd{\begin{document}} \def\ed{\end{document}}
\let\fr=\frac \let\bl=\bigl \let\br=\bigr
\let\Br=\Bigr \let\Bl=\Bigl
\let\bm=\bibitem
\let\na=\nabla
\def\tU{{\widetilde U}}
\let\pa=\partial \let\ov=\overline
\def\ie{{\it i.e.\ }}
\newcommand{\be}{\begin{equation}}
\newcommand{\ee}{\end{equation}}
\def\ba{\begin{array}}
\def\ea{\end{array}}
\def\ft#1#2{{\textstyle{{\scriptstyle #1}\over {\scriptstyle #2}}}}
\def\fft#1#2{{#1 \over #2}}
\def\F#1#2{{ F_{#1}^{(#2)} }}
\def\cF#1#2{{ {\cal F}_{#1}^{(#2)} }}

\def\R{{\bf R}}
\def\sst#1{{\scriptscriptstyle #1}}
\def\oneone{\rlap 1\mkern4mu{\rm l}}
\def\e7{E_{7(+7)}}
\def\td{\tilde}
\def\wtd{\widetilde}
\def\im{{\rm i}}
\def\bog{Bogomol'nyi\ }
\newcommand{\ho}[1]{$\, ^{#1}$}
\newcommand{\hoch}[1]{$\, ^{#1}$}
\newcommand{\bea}{\begin{eqnarray}}
\newcommand{\eea}{\end{eqnarray}}
\newcommand{\ra}{\rightarrow}
\newcommand{\lra}{\longrightarrow}
\newcommand{\Lra}{\Leftrightarrow}
\newcommand{\ap}{\alpha^\prime}
\newcommand{\bp}{\tilde \beta^\prime}
\newcommand{\cB}{{\cal B}}
\newcommand{\cO}{{\cal O}}
\newcommand{\vecx}{\vec{x}}
\newcommand{\vecy}{\vec{y}}
\newcommand{\vecp}{\vec{p}}
\newcommand{\vecq}{\vec{q}}
\newcommand{\NP}{Nucl. Phys. }

\newcommand{\cL}{{\cal L}}
\newcommand{\cA}{{\cal A}}
\newcommand{\cT}{{\cal T}}
\newcommand{\cR}{{\cal R}}
\newcommand{\cD}{{\cal D}}
\newcommand{\cH}{{\cal H}}

\newcommand{\SumInt}{\mathchoice
  {\rlap{$\displaystyle\sum$}\kern0.14em\int}
  {\rlap{$\textstyle\sum$}\kern0.14em\int}
  {\rlap{$\scriptstyle\sum$}\kern0.08em\int}
  {\rlap{$\scriptscriptstyle\sum$}\kern0.05em\int}
}

\def\Cb{\bar{C}}

\def\sst#1{{\scriptscriptstyle #1}}
\def\0{{\sst{(0)}}}
\def\1{{\sst{(1)}}}
\def\2{{\sst{(2)}}}
\def\3{{\sst{(3)}}}
\def\4{{\sst{(4)}}}
\def\5{{\sst{(5)}}}
\def\6{{\sst{(6)}}}
\def\7{{\sst{(7)}}}
\def\8{{\sst{(8)}}}
\def\9{{\sst{(9)}}}
\def\p{{\sst{(p)}}}
\def\q{{\sst{(q)}}}
\def\ve{\varepsilon}
\def\vf{\varphi}
\def\wg{\wedge}

\def\e{\epsilon}
\def\barl{\bar{l}}

\def \bi{\bibitem}
\def \la {\label}

\def \l {\lambda}
\def\foot{\footnote}
\def \tl  {{\tilde \l}}
\def \sql {{\sqrt \l}}
\def \adss {$AdS_5 \times S^5$\ }
\newcommand{\rf}[1]{(\ref{#1})}
\def \ov {\over}

\def\th{\theta}
\def\Th{\Theta}
\def\vth{\vartheta}
\def\btheta{{\bar\theta}}
\def\ttheta{{{\tilde\theta}}}
\def\bttheta{{{\bar\ttheta}}}
\def\vth{\vartheta}

\def\ra{\rightarrow}
\def\N{\nabla}
\def\F{{\cal F}}
\def\uM{\underline{M}}
\def\uA{\underline{A}}
\def\uN{\underline{N}}
\def\uP{\underline{P}}
\def\ua{\underline{a}}
\def\ub{\underline{b}}
\def\uc{\underline{c}}
\def\ud{\underline{d}}
\def\ue{\underline{e}}
\def\uf{\underline{f}}
\def\ui{\underline{i}}
\def\uj{\underline{j}}
\def\uk{\underline{k}}
\def\ul{\underline{l}}
\def\ual{\underline{\alpha}}
\def\ube{\underline{\beta}}
\def\um{\underline{m}}
\def\un{\underline{n}}
\def\up{\underline{p}}
\def\uq{\underline{q}}
\def\ur{\underline{r}}
\def\us{\underline{s}}
\def\umu{\underline{\mu}}
\def\unu{\underline{\nu}}
\def\ula{\underline{\l}}
\def\uka{\underline{\k}}
\def\usi{\underline{\s}}
\def\urh{\underline{\r}}
\def\cc{\circ}
\def\eqv{\equiv}

\def\ni{\noindent}

\def\Ep{E^{{}^{(+)}}}
\def\Em{E^{{}^{(-)}}}

\def\Mp{M^{{}^{(+)}}}
\def\Mm{M^{{}^{(-)}}}

\def \ha{{1\ov 2}}

\def\r{\rho}

\def\Y{{\rm Y}}
\def\X{{\rm X}}
\def\tY{\tilde{\rm Y}}
\def\tX{\tilde{\rm X}}
\def\dY{\dot{\rm Y}}
\def\dX{\dot{\rm X}}

\def \J {\mathcal{J}}
\def \del {\partial}

\def\dF{\dot{F}}
\def\dG{\dot{G}}
\def\df{\dot{f}}
\def \E {{\cal E}}
\def \S {{\cal S}}
\def \J {{\cal J}}

\def\ms{\mathcal{S}}
\def\mj{\mathcal{J}}
\def\soj{\fr{\ms}{\mj}}
\def \R {{\bf R}}
\def \om {\omega}
\def \bE {\bar E}
\def \x {{\cal X}}

\def \bi{\bibitem}
\def \la {\label}

\def \l {\lambda}
\def\foot{\footnote}
\def \tl  {{\tilde \l}}
\def \sql {{\sqrt \l}}
\def \adss {$AdS_5 \times S^5$\ }
\def \ov {\over}

\def \varpi {{\rm w}}

\def\thb{\bar{\theta}}
\def\Thb{\bar{\Theta}}
\def\barp{\bar{p}}
\def\barq{\bar{q}}
\def\barc{\bar{c}}
\def\bard{\bar{d}}
\def\bare{\bar{e}}

\def\thb{\bar{\theta}}
\def\Thb{\bar{\Theta}}
\def\mb{\bar{\m}}
\def\ab{\bar{\a}}
\def\zb{\bar{z}}
\def\psib{\bar{\psi}}
\def\barl{\bar{l}}
\def\barp{\bar{p}}
\def\barq{\bar{q}}
\def\barc{\bar{c}}
\def\bard{\bar{d}}
\def\baru{\bar{u}}

\def\e{\epsilon}
\def\wb{\bar{w}}
\def\lb{\bar{\l}}
\def\Jb{\bar{J}}
\def\Nb{\bar{N}}
\def\Zb{\bar{Z}}
\def\pab{\bar{\pa}}

\def\At{\tilde{A}}
\def\Bt{\tilde{B}}
\def\Ct{\tilde{C}}
\def\Dt{\tilde{D}}
\def\Et{\tilde{E}}
\def\Ft{\tilde{F}}
\def\Gt{\tilde{G}}
\def\Ht{\tilde{H}}
\def\Kt{\tilde{K}}
\def\Mt{\tilde{M}}
\def\Nt{\tilde{N}}
\def\Rt{\tilde{R}}
\def\at{\tilde{a}}
\def\bt{\tilde{b}}
\def\ct{\tilde{c}}
\def\dt{\tilde{d}}
\def\et{\tilde{e}}
\def\ft{\tilde{f}}
\def \zt{\tilde{z}}
\def \ztt{\tilde{\z}}
\def\Omt{\tilde{\Omega}}
\def \zetat{\tilde{\zeta}}
\def\htil{\tilde{h}}
\def\gt{\tilde{g}}
\def\nt{\tilde{n}}
\def\mut{\tilde{\mu}}
\def\nut{\tilde{\nu}}
\def\pht{\tilde{\f}}
\def\Phit{\tilde{\Phi}}
\def\vft{\tilde{\vf}}
\def\etat{\tilde{\eta}}

\def\rht{\tilde{\rho}}

\def\asth{\hat{*}}
\def\phh{\hat{\phi}}

\def\bA{{\bf A}}

\def\ola{\overleftarrow}
\def\ora{\overrightarrow}
\def\alt{\tilde{\a}}

\def\eh{\hat{e}}
\def\eph{\hat{\e}}
\def\ph{\hat{p}}
\def\alh{\hat{\a}}
\def\beh{\hat{\b}}
\def\gah{\hat{\g}}
\def\Fh{\hat{F}}
\def\muh{\hat{\m}}
\def\nuh{\hat{\n}}
\def\thh{\hat{\th}}
\def\rhh{\hat{\r}}
\def\dh{\hat{d}}
\def\ih{\hat{i}}
\def\jh{\hat{j}}
\def\hh{\hat{h}}
\def\nh{\hat{n}}
\def\gh{\hat{g}}
\def\kh{\hat{k}}
\def\deh{\hat{\d}}
\def\wh{\hat{w}}
\def\lah{\hat{\l}}
\def\Ah{\hat{A}}
\def\Gh{\hat{G}}
\def\Kh{\hat{K}}
\def\Nh{\hat{N}}
\def\Rh{\hat{R}}
\def\Ch{\hat{C}}
\def\Omh{\hat{\Omega}}

\def\xh{\hat{x}}

\def\ps{\rlap{\, /}\;\,p }
\def\ks{\rlap{\, /}\;\,k }

\def\gym{g_{YM}}

\def\adot{\dot{a}}
\def\bdot{\dot{b}}
\def\bpa{\bar{\pa}}

\def\pr{\prime}
\def\ssk{\medskip}
\def\clb{\color{blue}}
\def\clr{\color{red}}
\def\clg{\color{green}}
\def\clp{\color{purple}}
\def\clc{\color{cyan}}
\def\clm{\color{magenta}}
\def\cly{\color{yellow}}
\def\ah{\color{orange}}

\def\bfA{{\bf A}}
\def\bfB{{\bf B}}
\def\bfK{{\bf K}}
\def\bfU{{\bf U}}
\def\bfX{{\bf X}}
\def\bfY{{\bf Y}}
\def\bfZ{{\bf Z}}
\def\bfg{{\bf g}}
\def\bfn{{\bf n}}

\def\bsk{\bigskip}
\def\ssk{\medskip}

\def\Ec{{\cal E}}

\vspace{ -3cm}
 \vspace{0.1cm}

\title{ Stochastic analysis of finite-temperature effects on cosmological parameters}
 \date{}

\author[1]{Armin Hatefi\footnote{ahatefi@mun.ca}}
\author[2,3]{Ehsan Hatefi\footnote{ehsanhatefi@gmail.com}}
\author[4]{I. Y. Park\footnote{inyongpark05@gmail.com}}
\affil[1]{\small{Department of Mathematics and Statistics, Memorial University of Newfoundland, St. John’s, NL, Canada.}}
\affil[2]{\small{University of Alcala, Department of Signal Theory and Communications, Madrid, Spain}}
\affil[3]{\small{Scuola Normale Superiore and I.N.F.N,
Piazza dei Cavalieri 7, 56126, Pisa, Italy}}
\affil[4]{\small{Department of Applied Mathematics, Philander Smith University, Little Rock, AR 72202, USA.}}

\begin{document}
 \maketitle

\begin{abstract}

We explore the impact of finite-temperature quantum gravity effects on
cosmological parameters, particularly the effective vacuum-energy sector, by
incorporating temperature-dependent quantum corrections into the Hubble
parameter. To this end, we modify the Cosmic Linear Anisotropy Solving System
and introduce new density parameters, $\Omega_{\Lambda_2}$ and
$\Omega_{\Lambda_3}$, arising from finite-temperature quantum gravity
contributions. These parameters encode off-shell vacuum-bubble contributions
from Standard Model fields --- including heavy fields that cannot be treated
as on-shell radiation --- and may equivalently be viewed as quantum
corrections to the effective equation of state of the early-Universe plasma.
We analyze their influence on the cosmic microwave background power spectrum
using machine learning techniques, including artificial neural networks and
stochastic optimization. Our results reveal that $\Omega_{\Lambda_2}$ assumes
a negative value, consistent with dimensional regularization in
renormalization, and that the inclusion of $\Omega_{\Lambda_2}$ and
$\Omega_{\Lambda_3}$ improves the fit to 2018 Planck data. The present
approach is a phenomenological model-building exercise: a first-principles
derivation that would rigorously separate on-shell kinetic-theory content
from off-shell vacuum contributions has not yet been carried out, but the
approach captures the leading effects. Although the Hubble tension persists,
our findings highlight the potential of quantum gravitational corrections in
refining cosmological models and motivate further investigation into
higher-order thermal effects and polarization data constraints.

\end{abstract}

\newpage

\section{Introduction }

The cosmological constant problem remains one of the most profound challenges in theoretical cosmology (see \cite{Sola:2013gha, SolaPeracaula:2022hpd, SolaPeracaula:2026trz} for reviews and recent developments), with the observed value of vacuum energy being many orders of magnitude smaller than quantum field theory predictions. Recent works \cite{Park:2021ohu,Park:2021vro,Park:2024kfn}  employing finite-temperature quantum field theory (finite-T QFT) (see \cite{Kapusta,Bellac,Laine:2016hma} for reviews) suggest that temperature-dependent quantum corrections may provide new insights into this longstanding puzzle.\footnote{The central idea in \cite{Park:2021ohu,Park:2021vro} is the introduction of a novel renormalization scheme in which the renormalized masses are chosen to be of the order of the temperature. Similar approaches were adopted in earlier works, such as \cite{Moreno-Pulido:2020anb}, where problematic $m^4$ terms were eliminated through an alternative renormalization prescription.} These effects, particularly in the context of quantum gravity (QG), introduce perturbative corrections to the effective vacuum-energy sector that could significantly impact our understanding of early Universe physics.

In conventional cosmology (see \cite{Weinberg,Dodelson} for reviews), temperature effects are typically treated through classical thermodynamics, while quantum contributions are disregarded. However, as demonstrated in \cite{Park:2021ohu,Park:2021vro}, where it was proposed that the cosmological constant (CC) problem may be a consequence of how perturbation theory and finite renormalization are handled (see \cite{Sola:2013gha, SolaPeracaula:2022hpd}, \cite{Balazs:2022anb}, and \cite{Ageeva:2024qie} for related discussions), finite-T QFT effects naturally generate temperature-dependent contributions to the effective vacuum-energy sector in FLRW cosmology through loop-generated corrections. These contributions arise from off-shell vacuum-bubble diagrams of Standard Model fields - including heavy fields that are not captured by the standard on-shell radiation density $\Omega_r$ - and may equivalently be interpreted as quantum corrections to the equation of state of the early-Universe plasma (see section 2.1 for a detailed discussion of this dual
interpretation). This arises from the temperature dependence of the one-particle
irreducible (1PI) effective action, where the effective
cosmological-constant sector takes the form
$\Lambda_{\rm tot} = \Lambda_1 + \Lambda_2 a^{-4}
+ \Lambda_3 a^{-2} + \cdots$ with $a$ the scale factor,
introducing new density parameters $\Omega_{\Lambda_2}$ and
$\Omega_{\Lambda_3}$ beyond the standard $\Omega_{\Lambda_1}$.
These parameters are spatially homogeneous and enter only the
background expansion rate $H(t)$, rather than sourcing
perturbations in the way that on-shell radiation or spatial
curvature does.

Continuing the effort initiated in \cite{Park:2025iog}, this work presents an intensive investigation of these finite-temperature QG effects on the cosmological parameters by conducting the following tasks: firstly, we modify the Cosmic Linear Anisotropy Solving System (CLASS) Boltzmann code \cite{Blas:2011rf,CLASS home} to incorporate temperature-dependent corrections to the Hubble parameter, including the new QG-induced parameters $\Omega_{\Lambda_2}$ and $\Omega_{\Lambda_3}$. Secondly, we carry out an extensive numerical analysis by using both brute-force parameter scans and advanced machine learning techniques to constrain these parameters against the 2018 Planck data \cite{Planck:2018vyg}. Thirdly, we develop artificial neural network (ANN) methods to efficiently explore the high-dimensional parameter space and identify optimal cosmological configurations. Studies employing ANNs to investigate cosmological parameters include \cite{Chantada:2022bdf,Agarwal:2012ew,Pal:2023uje,Olvera:2021jlq,Gomez-Vargas:2022bsm}.

The inclusion of finite-T QG effects addresses several key
theoretical questions. First, the negative sign that emerges for
$\Omega_{\Lambda_2}$ in our analysis finds natural explanation
through dimensional regularization in the renormalization
procedure. Second, while these effects are small in the present
era, they become significant during the radiation-dominated epoch
and near recombination, potentially influencing the interpretation
of early dark energy \cite{Poulin:2018cxd}. Third, our
phenomenological approach provides a framework to test whether
quantum gravitational corrections could alleviate persistent
tensions like the Hubble tension \cite{Riess:2016jrr}. We
emphasize that the present analysis is a phenomenological
model-building exercise: a fully first-principles derivation of
the hydrodynamic system from the underlying finite-temperature
quantum field theory - analogous in spirit to Weinberg's
external-field treatment of QED \cite{Weinberg1} but extended to
the full Standard Model (SM) coupled to gravity at finite temperature
(see section~2.1 for details) - has not yet been carried out.
The simplified approach adopted here, in which the 1PI effective
action of the Standard Model+gravity system is coupled to a separately
specified hydrodynamic sector, captures the leading effects and
serves as a practical starting point.

Our methodology combines theoretical considerations - grounded
in the 1PI effective action but implemented via the simplified
coupling to a hydrodynamic sector described above - with
state-of-the-art computational techniques. The modified CLASS implementation \cite{mod_CLASS} allows us to systematically study how finite-T corrections affect the CMB power spectrum, while our machine learning (ML) approach enables efficient exploration of the eight-dimensional parameter space ($\Omega_{\Lambda_2}$, $\Omega_{\Lambda_3}$, $h$, $\omega_b$, $\omega_{cdm}$, $A_s$, $n_s$, $\tau_{reio}$).

\vspace{.2in}

The paper is organized as follows. Section 2 describes our modifications to CLASS. A natural question arises: given their similar scaling behaviors, could $\Omega_{\L_2}$ and $\Omega_{\L_3}$ be absorbed into existing parameters such as the radiation density $\Omega_r$ and the curvature density $\Omega_K$, respectively? We address this question. Section 3 outlines our numerical methods and results, including parameter estimation using artificial neural networks (ANNs). Our approach, which is a model-building exercise, differs from the standard Bayesian parameter inference, adopting a frequentist framework augmented by machine learning techniques; the rationale behind this choice is also discussed. It is noteworthy that both the results obtained in Table~\ref{tab:mse_metric_feature_removal} and Fig.~\ref{fig:mse_5k} indicate that $\Omega_{\L_2}$ and $\Omega_{\L_3}$ play a crucial role in determining the fine structure of the CMB power spectrum curve. In section 4, we conclude with a discussion of the implications for cosmological models and potential future directions. In Appendix~A, the non-degeneracy of $\Omega_{\L_3}$ and $\Omega_{K}$ is explicitly demonstrated. Appendix~B presents the ANN-based statistical method employed in the main analysis of section 3.

\section{Computational framework}

In this section, we outline the main computational tool developed and employed in this study to incorporate finite-T effects and analyze their cosmological implications, a modified version of CLASS Boltzmann code. We describe the modifications made to the CLASS to account for finite-temperature quantum gravitational corrections, which manifest as shifts in cosmological parameters, most notably the cosmological constant. These modifications enable the computation of power spectra under the influence of temperature-dependent vacuum energy contributions.

\subsection{Conceptual framework}
\label{sec:conceptual}

Before delving into the specific computational tools used in this
work, it is useful to outline a more conceptual framework that
motivates our overall approach, and to clarify in what sense the
newly introduced parameters $\Omega_{\Lambda_2}$ and
$\Omega_{\Lambda_3}$ encode physical effects that are not already
accounted for by the standard density parameters $\Omega_r$ and
$\Omega_K$ of $\Lambda$CDM cosmology.

\paragraph{1PI effective action and two equivalent bookkeepings.}
At a fundamental level, the starting point would be the action
describing the Standard Model (SM) fields coupled to gravity,
supplemented by a macroscopic matter system modeling the large-scale
thermodynamic evolution of the Universe.\footnote{In this section, we consider various systems to explain the rationale behind our framework. The actual system analyzed in section 3 is essentially the same as that considered by CLASS, except for the shifted CC.} Since quantum effects are
expected to play a significant role - particularly in the early
Universe - it is natural to consider the one-particle irreducible
(1PI) effective action for the coupled system. More specifically,
one can take the classical action of the form
\begin{equation}
S = S_{\rm EH} + S_{\rm SM} + S_{\rm PF}
\label{eq:totalaction}
\end{equation}
where $S_{\rm EH}$ denotes an Einstein--Hilbert action with a
cosmological constant, $S_{\rm SM}$ the SM action (in a curved
background), and $S_{\rm PF}$ a hydrodynamic or kinetic-theory
matter system - say, a perfect fluid, to be specific:
\begin{align}
\mathcal{L}_{\rm EH} &= \frac{1}{\kappa^2}(R - 2\Lambda)
\nonumber\\[4pt]
\mathcal{L}_{\rm SM} &= \mathcal{L}_{\rm gauge+gh}
+ \mathcal{L}_{\rm fer} + \mathcal{L}_{\rm Yukawa}
+ \mathcal{L}_{\rm Higgs}
\label{eq:LSM}
\end{align}
where the contractions of the spacetime indices are done with the
curved metric; the covariant derivatives contain both the Christoffel
and gauge connections. The detailed forms and field contents of each
line above as well as the overall conventions can be found in
\cite{Park:2024kfn}. The explicit form of the action $S_{\rm PF}$
in the hydrodynamic matter case can be found, e.g., in
\cite{Brown:1992kc}. One can quantize the system
\cite{Park:2024kfn}, and obtain the 1PI action. For cosmological
applications, the important terms will be the classical part (or
renormalized part, more precisely) and leading terms in the
derivative expansion, as higher-order contributions are typically
suppressed. More precisely, one can focus on the renormalized action
with the coupling constants shifted by quantum effects.\footnote{As
a slightly different and conceptually simpler approach - yet one
that ultimately leads to the same starting point at the level of
approximation currently adopted for cosmological studies as just
described - one may employ the following setup. In this approach,
one first constructs an effective field theory action without
incorporating the kinetic theory component. The term ``effective''
can carry different meanings in the literature; here, it refers
specifically to an effective field theory obtained by retaining only
the leading terms in the derivative expansion of the 1PI action.
Notably, quantum corrections to the cosmological constant are
included. The kinetic theory system can then be coupled to this
resulting field-theoretic system. \la{fn3}}

\begin{figure}[t]
\begin{center}
\begin{fmffile}{vacandtadcor}
\parbox{40mm}{
  \begin{fmfgraph*}(75,50)
     \fmfi{gluon}{reverse fullcircle scaled .5w shifted (.5w,.5h)}
  \end{fmfgraph*}
}  
\parbox{40mm}{
  \begin{fmfgraph*}(75,50)
     \fmfi{dashes}{reverse fullcircle scaled .5w shifted (.5w,.5h)}
  \end{fmfgraph*}
}  
\end{fmffile}
 \end{center} 
\[\hspace{-.5in}\mbox{(a)}\hspace{1.4in} \mbox{(b)}\] 
       \caption{vacuum diagrams in the pure gravity sector: (a) graviton loop (b) ghost loop}
\label{vtvt}
\end{figure}
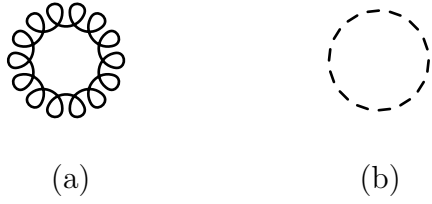

Especially, non-derivative quantum corrections, such as those
contributing to the cosmological constant, can have a non-negligible
impact despite their small magnitude. This is so because the observed
value of the cosmological constant is itself extremely small, making
even suppressed quantum effects potentially significant. Similar
roles for non-derivative quantum corrections have been explored in
other contexts, such as black hole accretion physics
\cite{Nurmagambetov:2020ann}, underscoring their physical relevance
in some situations. Shifts in the cosmological constant arise from
the vacuum-to-vacuum diagrams. Fig.~\ref{vtvt} shows the relevant diagrams
in the pure gravity sector (analogous diagrams exist in the
matter-coupled sector). For example, the graviton
vacuum-to-vacuum amplitude is given by:
\begin{equation}
\int \prod_x dh_{\kappa_1\kappa_2}\;
e^{\frac{i}{\kappa'^{\,2}}\int\sqrt{-\tilde{g}}\,
\bigl(-\tfrac{1}{2}\tilde{\nabla}_\gamma h_{\alpha\beta}\,
\tilde{\nabla}^\gamma h^{\alpha\beta}\bigr)}\;.
\label{eq:gravitonvac}
\end{equation}
Evaluation of such diagrams including the ones coming from the SM
sector in a finite temperature, an example of which is given below,
generically leads to the form of the CC given in the introduction
\cite{Park:2021ohu,Park:2021vro,Park:2024kfn}. For relevance of
dynamic vacuum in the context of the CC problem, see
\cite{Moreno-Pulido:2022phq,Moreno-Pulido:2023ryo}.

A noteworthy observation in this work is that at the level of the 1PI action, the
finite-temperature quantum corrections generated by SM fields (and
gravitons) admit \emph{two equivalent descriptions}:
\begin{enumerate}
\item[(i)] as temperature-dependent shifts to the cosmological
constant sector of the effective action, leading to a time-dependent
effective $\Lambda$, or
\item[(ii)] as improved modelling of the effective equation of state
of the early-Universe plasma at the quantum level, arising from
vacuum-bubble contributions of all SM fields --- including those
that cannot be treated as on-shell radiation.
\end{enumerate}
These two interpretations are related by bookkeeping within the 1PI
action: the quantity $J(m_\phi, T)$ appearing in
Eq.~\eqref{eq:Jdef} below is simultaneously the Helmholtz free
energy density of the quantized field and (when extracted as a
$\phi$-independent, $\sqrt{-g}$-multiplied term) a
temperature-dependent contribution to the effective
cosmological-constant sector. The phenomenological analysis carried
out in this work does not depend on which interpretation one favours;
what matters for the Hubble parameter, and hence for the CMB power
spectrum, is the presence of the resulting $T^4$- and $T^2$-scaling
contributions, which are parametrized by $\Omega_{\Lambda_2}$ and
$\Omega_{\Lambda_3}$.

\paragraph{Why these contributions are not already in $\Omega_r$.}
A natural concern is whether the $T^4$ contribution associated with
$\Omega_{\Lambda_2}$ is simply the pressure of a relativistic gas,
and therefore already included in the standard radiation density
$\Omega_r$ on the right-hand side of the Friedmann equation. This
concern is well-motivated and deserves a careful answer.

The first term on the right-hand side of Eq.~\eqref{eq:Jexp} below,
$-\frac{\pi^2}{90}T^4$, is the one-loop thermal free energy density
of a non-interacting massless scalar (equivalently, $f = -p$ for a
relativistic gas). However, for one thing, the finite-temperature 1PI analysis of the full SM
\cite{Park:2024kfn} produces $T^4$ and $T^2$ contributions from
\emph{every} field in the theory, including fields that are
\emph{not} relativistic at the temperatures of interest and therefore
cannot be assigned to $\Omega_r$. Concretely, the Higgs boson, the
top quark, and the electroweak gauge bosons have masses far above
the recombination-era temperature; they do not propagate as on-shell
radiation, do not appear in the kinetic-theory description of the
plasma, and are consequently absent from the standard $\Omega_r$.
Their contributions to the Hubble parameter arise through
vacuum-bubble diagrams in the 1PI effective action - i.e.\ as
virtual particles - and these contributions are \emph{not}
captured by the standard $\Lambda$CDM bookkeeping. It is such contributions, summed over the SM field content (and in
principle the graviton), that $\Omega_{\Lambda_2}$ and
$\Omega_{\Lambda_3}$ parametrize.

Equivalently, in the EOS-modification language of interpretation
(ii) above: the effective equation of state of the early-Universe
plasma at the quantum level differs from the naive kinetic-theory
form $p = \rho/3$ (for radiation) plus $p = 0$ (for dust), precisely
because heavy virtual fields contribute to the vacuum energy in a
temperature-dependent way. The parameters $\Omega_{\Lambda_2}$ and
$\Omega_{\Lambda_3}$ quantify this departure at leading order in the
high-temperature expansion. Higher-order corrections, such as those
associated with Debye screening, collective excitations, and other
in-medium effects, would further refine this EOS modelling and
generate contributions with non-$T^4$ scalings; we do not include
them in the present work but regard them as natural targets for
future investigation.

\paragraph{First-principles derivation and the approach adopted here.}
The first-principles approach to rigorously establishing the
separation between on-shell kinetic-theory content and off-shell
vacuum contributions would proceed as follows. One starts with the
SM coupled to gravity and quantizes the system \`a la Weinberg's
treatment of the external-field approximation in QED
(Ch.~13.6 of \cite{Weinberg1}), where a $U(1)$ gauge system was
considered in the presence of a heavy charged particle. In that
analysis, the relevant diagrams are the ladder-type graphs (Fig.~13.5
of \cite{Weinberg1}) in which photon lines connect the light
electron line to the heavy-particle line. Weinberg shows that,
in the limit of a slowly moving heavy particle, the sum of all such
ladder diagrams factorizes and is equivalent to a set of Feynman
rules in which the light electron interacts with a classical
external vector potential $\mathcal{A}^\mu(x)$ (see Eq.~(13.6.15) of \cite{Weinberg1}) - equivalently, to
adding a new interaction term
$\mathcal{L}_{\rm ext}(x) = \mathcal{A}_\mu(x)\,J_e^\mu(x)$
(Eq.~(13.6.14) of \cite{Weinberg1}) to the Lagrangian where
\bea
J_e^\mu(x)\equiv -ie \bar{\Psi} \g^\m \Psi
\eea 
is the electric current of the electrons. The heavy
particle's quantum-level
interaction with the light electron through the ladder diagrams is
repackaged as an effective external-field interaction. Crucially,
vacuum bubble diagrams - which lead to a shifted CC - are topologically distinct from these ladder
graphs in that they do not connect to the heavy-particle line.
They therefore contribute separately to the effective action, viz, to the CC sector.

An analogous procedure is required in the present case in order to
derive the hydrodynamic system from first principles, in particular, the source terms such as $\r_r$ and $\r_m$ in the Lagrangian, just as $\mathcal{L}_{\rm ext}(x) = \mathcal{A}_\mu(x)\,J_e^\mu(x)$ was derived. However, it is
far more complicated for several reasons. Firstly, one must extend
Weinberg's analysis to a finite-temperature setup. Secondly, the
system involves the full SM field content, and one must carry out
spontaneous symmetry breaking in a curved background (before honing
in on, say, the $U(1)$ sector for the derivation of the $\r_r$ term). The analogue of Weinberg's ladder-diagram
analysis, now applied to the relevant sectors of the SM + gravity in the presence
of radiation and dust (or whatever hydrodynamic components are
relevant), is expected to produce the usual hydrodynamic part - the
$\rho_r$ term in particular emerging as the analogue of
Eq.~(13.6.14). At the same time, vacuum bubble diagrams - including
those of the $U(1)$ sector itself, as well as those of the other SM
sectors - are topologically distinct from the ladder diagrams and
therefore contribute separately to the CC part of the effective
action. In the SM+gravity case, these vacuum-bubble contributions
are what $\Omega_{\Lambda_2}$ and $\Omega_{\Lambda_3}$ parametrize
at leading order. The end result is an effective action in which
the hydrodynamic part (carrying $\rho_r$, $\rho_{\rm cdm}$, etc.)
and the shifted CC part (carrying the $T^4$ and $T^2$ vacuum-bubble
contributions) are generated by distinct classes of diagrams, and
are therefore not in conflict: adding them does not double-count
the $U(1)$ sector, precisely because ladder-type and vacuum-bubble
diagrams are disjoint topological classes.

This procedure, while prohibitively complicated, can in principle be
carried out; it amounts to a concrete implementation of the
statement made in \cite{Park:2025iog} that ``ideally, a
second-quantized description whose effective description is provided
by the hydrodynamics should be employed.'' The approach actually
adopted in the present work (see also footnote~\ref{fn3}) is not the same as
this first-principles derivation. Instead, we take a simplified view
in which the 1PI action of the SM+gravity system is coupled to a
separately specified hydrodynamic sector. Strictly speaking, this is
not the same as the first-principles derivation of the effective
action just described. But since we are interested in the leading
terms for a model-building exercise, it serves as a practical
starting point, with the topological separation of ladder and
vacuum-bubble diagrams in the Weinberg analogue providing the
conceptual justification for why the on-shell hydrodynamic content
and the off-shell vacuum-bubble content can be treated as distinct.

\paragraph{One-loop generation in a simple setting.}
To avoid unnecessary complications associated with gravity
quantization, for the moment let us consider a scalar system to
illustrate one-loop generation of the CC. In a flat background:
\begin{equation}
S_{\rm ren} = -\int d^4x\,\Big[\tfrac{1}{2}\partial_\mu\phi\,
\partial^\mu\phi + \tfrac{1}{2}m^2\phi^2
+ \tfrac{\lambda}{4!}\phi^4\Big]\;.
\label{eq:Sscalar}
\end{equation}
Define
\begin{equation}
V_{\rm ren} \equiv \tfrac{1}{2}m^2\phi^2
+ \tfrac{\lambda}{4!}\phi^4\;,\qquad
m_\phi^2 \equiv m^2 + \tfrac{1}{2}\lambda\phi^2\;.
\label{eq:Vren}
\end{equation}
The one-loop effective potential at finite temperature is given by
(see e.g., \cite{Arnold:1992rz}, the work well-known in the
finite-$T$ literature, as well as \cite{Laine:2016hma})
\begin{equation}
V_{\rm one\text{-}loop} = V_{\rm ren} + J(m_\phi, T)
\label{eq:Jdef}
\end{equation}
\begin{align}
J(m_\phi, T) &\equiv \SumInt \tfrac{1}{2}
\ln\bigl({\clp P}^2 + m_\phi^2\bigr)
\nonumber\\
&= -\frac{\pi^2}{90}T^4 + \frac{m_\phi^2}{24}T^2
- \frac{1}{12\pi}m_\phi^3\,T
\nonumber\\
&\quad
- \frac{\mu^{-2\epsilon}}{2(4\pi)^2}\,m_\phi^4
\Bigl[\ln\!\Bigl(\frac{\bar\mu\,e^\gamma}{4\pi T}\Bigr)
+ \frac{1}{2\epsilon}\Bigr]
+ \frac{\zeta(3)}{3(4\pi)^4}\,\frac{m_\phi^6}{T^2}
+ \mathcal{O}\!\Bigl(\frac{m_\phi^8}{T^4}\Bigr)
+ \mathcal{O}(\epsilon)
\label{eq:Jexp}
\end{align}
where {\clp $P=(\omega_n,{\bf p})$ with $\omega_n$ the Matsubara frequencies and $\bf{p}$ the spatial loop momentum}; $\ln\bar\mu^2 \equiv \ln(4\pi\mu^2) - \gamma$ with $\mu$
being the scale parameter of dimensional regularization; $\gamma$ is
an Euler's constant; $\epsilon$ is a regularization parameter of
dimensional regularization; $\zeta(3)$ is the zeta function evaluated
at 3. (The above equation relies on a high-temperature expansion. Whether it remains valid to consider this approach over its low-temperature counterpart when the temperature is not sufficiently high compared to the particle mass is a subtle issue, which we discuss in the conclusion.) Before proceeding, we pause to fix the terminology that is central to this
work, since the words ``on-shell'' and ``off-shell'' will be used in a
specific sense throughout. By \emph{on-shell} we mean that a solution of the
field equation has been imposed: in the present scalar example, the trivial
solution $\phi=0$ (more generally, a stationary point of the effective
action). By \emph{off-shell} we mean that no such solution has been
substituted, so that the field $\phi$ is kept arbitrary and the
$\phi$-dependence of the effective potential is retained. We distinguish this term from momentum space mass-shell condition. The 1PI
effective action can be constructed either through the canonical operator formalism or through the path integral, and in the latter one need not place
the external lines on the momentum mass shell. The same flexibility persists
at finite temperature. (It is also worth
recalling that this freedom, i.e., on-mass-shell operator formalism vs. offshell path integral, is already present at zero temperature.)

The reason this distinction matters here is that $J(m_\phi,T)$ in Eq.~(8) is
obtained, in the path-integral formalism, \emph{without} imposing the momentum mass-shell condition (or $\f=0$, of course);
it is therefore a genuinely off-shell object,\footnote{One subtlety is that fixing the temperature to that of the FLRW background corresponds to partial specification of the background, and in this regard, can be viewed as "partially onshell."} depending on the background field. When $m=0$, the derivation of $p=\tfrac{1}{3}\rho$, which allows for the massless field under consideration to be identified as radiation, in the quantum field theory framework requires the on-shell reduction $\phi=0$;
this is discussed in detail in the companion paper~\cite{Park:2025iog} (see also
the paragraph ``On-shell thermodynamics vs off-shell vacuum contributions''
below). The $\phi$-independent, temperature-dependent vacuum-bubble pieces are precisely the contributions we collect into the
effective cosmological-constant sector and parametrize, at leading orders, by
$\Omega_{\Lambda_2}$ and $\Omega_{\Lambda_3}$. For the contributions to the CC, one can note the
$\phi$-independent terms: the first two leading order terms scale
$\sim T^4$ and $\sim T^2$. To extend the discussion to an FLRW
background, let us extend the system to a curved background.
\begin{equation}
S_{\rm ren} = \frac{1}{\kappa^2}\int d^4x\,\sqrt{\gamma}\,
(R - 2\Lambda_1) - \int d^4x\,\sqrt{-g}\,
\bigl(\tfrac{1}{2}g^{\mu\nu}\partial_\mu\phi\,\partial_\nu\phi
+ V\bigr)
\label{eq:Scurved}
\end{equation}
where $\kappa^2 = 16\pi G$ with $G$ Newton's constant; $\kappa^2$
will be suppressed in some places. As is well known, an FLRW
background has a time-dependent temperature. The
standard finite-temperature techniques cannot in general be directly
applied to a system with a non-constant temperature. This issue has been recently addressed in
\cite{Park:2024kfn}: for an FLRW background, the standard finite-$T$
techniques can be applied with minor modifications. This is due to
the special feature of an FLRW background (say, the flat case in which we
are interested) that the temperature scales as $\sim\frac{1}{a}$ and
the scale factor happens to be a metric component:
\begin{equation}
ds^2 = -dt^2 + a^2\,(dx^2 + dy^2 + dz^2)
\label{eq:FLRWmetric}
\end{equation}
Therefore the temperature is annihilated by the covariant derivative
and the upshot is that the temperature $T$ appearing in the results
such as \eqref{eq:Jexp} can simply be replaced by the time-dependent
one $T(t)$. The result \eqref{eq:Jexp} thus implies
\begin{equation}
\Lambda \equiv \Lambda_1 + \frac{\Lambda_2}{a^4}
+ \frac{\Lambda_3}{a^2} + \cdots
\label{eq:Lambdatot}
\end{equation}
where $\Lambda_2$ and $\Lambda_3$ are numerical constants introduced
as part of the definition of $\Lambda$. $\Omega_{\Lambda_2}$ and
$\Omega_{\Lambda_3}$ are then introduced as the corresponding
density parameters, which implies the modified form of the Hubble
constant given in \eqref{eq:Hmod}. We emphasize once more that
Eq.~\eqref{eq:Lambdatot} is one convenient bookkeeping;
equivalently, these terms correspond to quantum modifications of the
effective plasma EOS not captured by standard kinetic-theory
treatments.

\paragraph{Non-degeneracy with $\Omega_K$.}
A natural question arises as to whether $\Omega_{\Lambda_2}$ and
$\Omega_{\Lambda_3}$ could be absorbed into existing parameters such
as the radiation density $\Omega_r$ and curvature density $\Omega_K$,
respectively, given their similar scaling behaviors. For
$\Omega_{\Lambda_2}$, the discussion above already clarifies the
situation: even though it scales as $a^{-4}$, it encodes
vacuum-bubble contributions that are absent from
$\Omega_r$, and therefore it cannot be absorbed into the latter.

For $\Omega_{\Lambda_3}$ vs $\Omega_K$, the non-degeneracy can be
seen most cleanly in the simplest possible setting. Consider the
Einstein--Hilbert action including a cosmological constant:
\begin{equation}
\mathcal{L} = \frac{1}{\kappa^2}(-2\Lambda_1 + R)
\label{eq:EHaction}
\end{equation}
We expand the metric $g_{\mu\nu}$ around a background solution
$g_{0\mu\nu}(K)$ such that
$g_{\mu\nu} \equiv h_{\mu\nu} + g_{0\mu\nu}(K)$, where the
background is given by:
\begin{equation}
ds^2 = a^2\Bigl(dx^2 + K\,\frac{(x\cdot dx)^2}{1 - Kx^2}\Bigr)
\label{eq:curvedFLRW}
\end{equation}
By substituting this expansion into the action and taking the
$h_{\mu\nu}$-variation, one obtains the field equations. In this
context, the background curvature parameter $K$ appears in multiple
locations within the resulting equations. This explains why the
density parameter $\Omega_K$, defined as
\begin{equation}
\Omega_K \equiv -\frac{K}{a_0^2 H_0^2}
\label{eq:OmegaK}
\end{equation}
enters the standard cosmological equations through several distinct
channels. In contrast, the parameter $\Omega_{\Lambda_3}$ has an
entirely different physical origin. When finite-temperature one-loop
corrections are incorporated, the cosmological constant $\Lambda_1$
is shifted schematically as:
\begin{equation}
\Lambda_1 + \Lambda_2\,\frac{1}{a^4}
+ \Lambda_3\,\frac{1}{a^2} + \cdots
\label{eq:CCshift}
\end{equation}
Unlike $\Omega_K$, the parameter $\Omega_{\Lambda_3}$ originates
exclusively from the cosmological-constant term of the off-shell
one-loop effective action. Following a standard linearized analysis
of gravity coupled with hydrodynamic matter, it enters the Hubble
parameter $H$.\footnote{The fluctuations $\delta\rho_{\Lambda_2}$
and $\delta\rho_{\Lambda_3}$ are neglected due to their suppressed
nature --- being both quantum in origin and small variations. Thus,
their influence is minimal, and $\Omega_{\Lambda_2}$ and
$\Omega_{\Lambda_3}$ effectively appear only within the modified
Hubble parameter $H$. (This statement holds at the level of
approximation adopted in the present work. At higher orders,
additional terms such as $\Lambda R_{\mu\nu}R^{\mu\nu}$ are expected
to appear.) In contrast, $\Omega_K$ and $\Omega_r$ directly
contribute to the fluctuation equations and play a significant role
in the evolution of perturbations. In this work, $\Omega_{\Lambda_2}$
and $\Omega_{\Lambda_3}$ are treated phenomenologically, and their
inclusion is justified by consistency with numerical results.
(Theoretically, these parameters can be computed from first
principles via the renormalization procedure. However, finite
renormalization freedom necessitates experimental inputs, i.e., renormalization condtions, like the
present one to fix their values anyway.)} While $\Omega_K$ (once kept) and
$\Omega_{\Lambda_3}$ appear only in the combination
$\Omega_{\Lambda_3} + \Omega_K$ within the expression for $H$, these
parameters are fundamentally non-redundant:
\begin{itemize}
\item \textbf{Origins:} One is geometric (curvature), while the
other stems from quantum corrections (loop effects).
\item \textbf{Appearances:} They enter the action and resulting
field equations through distinct mathematical terms.
\end{itemize}
The non-degeneracy of these parameters has been explicitly verified
in Appendix~A through numerical analysis using a modified version of
the CLASS code.

As for $\Omega_{\Lambda_2}$ again, radiation arises from physical, on-shell
particles (e.g., photons, neutrinos) through, say, kinetic theory
physics, whereas the contributions under consideration are due to
virtual particles in loop corrections. The finite-temperature effects
we consider are analogous to vacuum energy contributions induced by
loop effects in QED photon and electron fields. Radiation density,
on the other hand, pertains to the energy density of physical photons
and neutrinos. Therefore, the finite-temperature effects we consider
are not included in the conventional radiation density and must be
accounted for separately.

\paragraph{On-shell thermodynamics vs off-shell vacuum contributions.}
For a system more akin to one considered in CLASS, let us examine a system governed by Boltzmann kinetic
theory - such as a photon gas - interacting with a scalar field
(e.g., the Higgs field) in a curved spacetime background. The total
action of the system is given by
\begin{equation}
S = S_{\rm EH} + S_{\rm sf} + S_{\rm kin},
\label{eq:Ssystem}
\end{equation}
where $S_{\rm EH}$ represents the Einstein--Hilbert action with a
cosmological constant, $S_{\rm sf}$ is the scalar field action, and
$S_{\rm kin}$ accounts for the kinetic sector. While the explicit
form of $S_{\rm kin}$ is known in certain models (see, e.g.,
\cite{Schutz:1970my,Carter:1992im}) - such as that of an ideal gas
- it will not be needed for our current purposes. The associated
metric field equation contains the stress-energy tensor:
\begin{equation}
T^{\mu\nu} = \frac{1}{\sqrt{-g}}\sum_n
\delta^3(x_n(t) - x)\,\frac{p_n^\mu(t)\,p_n^\nu(t)}{E_n(t)}\;.
\label{eq:Tmunu}
\end{equation}
To avoid unnecessary complications, we assume a fixed,
non-dynamical background metric. Our primary interest lies in the
$T^4$-scaling behavior relevant
to $\Omega_{\Lambda_2}$. The conventional analysis focuses
exclusively on the classical thermodynamics of the kinetic system,
which leads to the well-known exact $T^4$ dependence of the energy
density - see, for example, footnote~1 in section 3.1 of
\cite{Weinberg}. In contrast, the focus of this work is on quantum
field-theoretic effects originating from the scalar field sector
$S_{\rm sf}$ (as well as from other Standard Model particles and the
graviton in the full description, which do not qualitatively alter
the conclusions) at finite temperature. Importantly, such scalar
fields cannot be treated as radiation due to their large masses;
instead, their contributions to the cosmological constant arise
through virtual processes, specifically vacuum bubble diagrams.

To be concrete, consider the scalar sector. The one-loop correction
to the classical (more precisely, renormalized) action takes the form
given in Eq.~\eqref{eq:Jexp}. This expression represents an infinite
series expansion in powers of temperature, with the leading
contribution scaling as $T^4$. Importantly, this term explicitly
shifts the classical cosmological constant (CC) and is distinct from
the $T^4$ contribution arising from the kinetic sector discussed
earlier.

\vspace{.2in}
To summarize the bookkeeping adopted throughout the rest of this
paper:
\begin{itemize}
\item On-shell relativistic species (photons, relativistic neutrinos)
and on-shell non-relativistic matter (cold dark matter, baryons) are
described by the kinetic-theory / hydrodynamic sector $S_{\rm PF}$,
and contribute to $\Omega_r$, $\Omega_{\rm cdm}$, and $\Omega_b$ as
usual.
\item Off-shell vacuum-bubble contributions from \emph{all} SM fields
(including heavy ones) and the graviton, generated by the 1PI
effective action at finite temperature, are collected into the
effective cosmological-constant sector and parametrized at leading
order by $\Omega_{\Lambda_1}$, $\Omega_{\Lambda_2}$, and
$\Omega_{\Lambda_3}$.
\end{itemize}
With this bookkeeping, the
new parameters $\Omega_{\Lambda_2}$ and $\Omega_{\Lambda_3}$ encode
genuinely new physical content relative to standard $\Lambda$CDM.

\subsection{Modification of CLASS}
\label{sec:classmod}

With the conceptual framework established in
Section~\ref{sec:conceptual}, we now turn to the implementation of
these ideas in CLASS. The key point that carries over -
independently of whether one interprets the finite-$T$ contributions
as shifts to the effective cosmological constant or as quantum
modifications of the early-Universe plasma EOS - is that the Hubble
parameter acquires new $T^4$- and $T^2$-scaling contributions from
the full SM field content, on top of the standard $\Omega_r$ and
$\Omega_K$ terms of $\Lambda$CDM. The task is therefore to modify the
Friedmann equation inside CLASS to include these contributions and to
examine their phenomenological consequences.

In its default configuration, CLASS evolves a cosmological model
derived from the action
\begin{equation}
S = S_{\rm CC} + S_{\rm EH} + S_{\rm kin},
\end{equation}
where $S_{\rm CC}$ denotes the cosmological constant term,
$S_{\rm EH}$ is the Einstein--Hilbert action, and $S_{\rm kin}$
collectively represents the kinetic theory terms for matter and
radiation. We modify CLASS to accommodate the extended system in
which the standard cosmological constant contribution, denoted by
$\Lambda_1$, is replaced by the effective quantity $\Lambda$ defined
in Eq.~\eqref{eq:Lambdatot}. Equivalently - and without any
change to the resulting code - this may be viewed as incorporating
quantum corrections to the effective plasma EOS not captured by
the kinetic-theory sector $S_{\rm kin}$.

The $\Lambda$CDM model (see \cite{Weinberg,Dodelson} for reviews)
assumes a spatially flat or nearly flat Universe, with a cosmological
constant representing dark energy. It successfully explains the
large-scale structure of the Universe, CMB observations, and galaxy
formation. The model involves six primary parameters (or seven if
the curvature density parameter $\Omega_K$ is included), which
govern the composition and evolution of the Universe. These
parameters are:
\begin{equation}
h,\; \omega_b,\; \omega_{\rm cdm},\; A_s,\; n_s,\; \tau_{\rm reio}.
\label{eq:LCDMparams}
\end{equation}
Their definitions and cosmological roles are summarized in Table \ref{cps}.
\begin{table}[h!]
\centering
\begin{tabular}{|c|p{11cm}|}
\hline
\textbf{Parameter} & \textbf{Description} \\
\hline
$h$ & The dimensionless Hubble parameter, which determines the current expansion rate of the Universe. Its value is crucial for estimating the age and size of the Universe. \\
\hline
$\omega_b$ & The baryon density parameter, representing the fraction of the Universe's energy density in the form of ordinary matter, "baryons". \\
\hline
$\omega_{cdm}$ & The cold dark matter density parameter, indicating the energy density contribution from non-relativistic, non-baryonic dark matter. \\
\hline
$A_s$ & The amplitude of the primordial scalar perturbations, which parameterizes the magnitude of the initial density fluctuations. \\
\hline
$n_s$ & The spectral index of the primordial perturbations, describing the scale dependence of the primordial power spectrum. \\
\hline
$\tau_{reio}$ & The optical depth to reionization, quantifying the degree of suppression of CMB anisotropies due to scattering by free electrons produced during reionization. \\
\hline
$\Omega_K$ & The curvature density parameter, accounting for the spatial curvature of the Universe. A value of $\Omega_K = 0$ implies a flat Universe, while positive or negative values suggest open or closed spatial geometry. \\
\hline
\end{tabular}
\caption{Summary of the cosmological parameters}
\label{cps}
\end{table}

The CLASS code \cite{Blas:2011rf,CLASS home} is a computational tool
used to solve the Boltzmann equations and generate the power spectra
of cosmological observables within the $\Lambda$CDM framework. Once
the cosmological parameters are specified, CLASS produces the
relevant observables accordingly. However, when considering QG
effects at finite temperatures, modifications to this standard
framework become necessary. These arise because the 1PI effective
action generates $T^4$ and $T^2$ contributions from SM fields (and
the graviton) that are not already captured by the standard
$\Omega_r$ and $\Omega_K$, as explained in detail in
Section~\ref{sec:conceptual}. At the level of approximation adopted
in the present work, the net effect is a shift in the Hubble
parameter; the effective cosmological-constant sector (or
equivalently the plasma EOS) becomes time-dependent through the
dependence of $\Lambda$ on the scale
factor.\footnote{These works extend earlier studies of
foliation-based gravity quantization at finite temperature
\cite{Park:2014tia,Park:2016zgt,Park:2014noa,Park:2015xoa,Park:2018vci}.
In particular, the gauge $K = 0$ (or more generally $K = K_0$),
where $K$ is the trace of the second fundamental form, plays a key
role \cite{Park:2015xoa}. This approach parallels Witten's recent
treatment in \cite{Witten:2022xxp}.} Among the cosmological
parameters, the most prominent shift is in the cosmological constant
$\Lambda$ (denoted as $\Lambda_1$ in the present context), which
becomes temperature-dependent due to quantum effects
\cite{Park:2021ohu,Park:2021vro,Park:2024kfn}.\footnote{The code
was obtained in collaboration with K.-T.~Cho and M.-S.~Lee.}

Referring to \cite{Park:2024kfn} and prior studies on the 1PI
effective action for the Standard Model coupled to gravity, we find
that the expression for the Hubble parameter $H(t)$ is modified.
The original expression, as adopted in \cite{Weinberg}, is
\begin{equation}
H = H_0\left[\Omega_M\left(\frac{T}{T_{\gamma 0}}\right)^{\!3}
+ \Omega_R\left(\frac{T}{T_{\gamma 0}}\right)^{\!4}
\right]^{1/2}
= 7.204\times 10^{-19}\,T^{3/2}
\sqrt{1.523\times 10^{-5}\,T + \Omega_M h^2}\;.
\label{eq:Hstd}
\end{equation}
where $T$ denotes the temperature. (In \cite{Weinberg}, both the
$\Omega_{\Lambda_1}$ and curvature density parameter are neglected
for their small contributions.) The modified CLASS implementation
\cite{mod_CLASS} includes finite-temperature corrections, yielding
the following form:
\begin{equation}
H = 7.204\times 10^{-19}\,T^{3/2}
\sqrt{1.523\times 10^{-5}\,T
+ \left(\frac{2.725}{T}\right)^{\!3}\Omega_{\Lambda_1}h^2
+ \Omega_M h^2
+ \frac{T}{2.725}\,\Omega_{\Lambda_2}h^2
+ \frac{2.725}{T}\,\Omega_{\Lambda_3}h^2}\;.
\label{eq:Hmod}
\end{equation}
In the approximation adopted in the present work, this is the only
location where $\Omega_{\Lambda_2}$ and $\Omega_{\Lambda_3}$ appear.
This is consistent with the bookkeeping described at the end of
Section~\ref{sec:conceptual}: the new parameters enter the
background expansion through the 1PI vacuum-energy sector as
spatially homogeneous, background-only contributions, while the
on-shell species (photons, neutrinos, baryons, CDM) continue to be
described by the standard kinetic-theory / hydrodynamic treatment
inside CLASS. The phenomenological content of the model can be assessed independently
of whether one describes Eq.~\eqref{eq:Hmod} as encoding a
time-dependent effective cosmological constant or a quantum
correction to the plasma EOS.


\section{Numerical analysis}\label{sec:num_stu}

Our analysis employs a modified version of CLASS incorporating the quantum gravity modifications previously described. The parameter space exploration proceeds in two distinct phases. First, we conduct a comprehensive brute-force scan across eight cosmological parameters spanning generous ranges to ensure complete coverage of the parameter space. By
these comprehensive scans tens of millions of data points are obtained. Following this initial broad exploration, we implement a refined search strategy focusing on regions that minimize the Euclidean distance (see the comments below) between our computed power spectra and the Planck dataset. This targeted approach yields significantly improved precision, with the minimal achieved distance of 26.83 corresponding to parameter values largely consistent with existing literature.\footnote{To give a sense of the accuracy associated with this number, the distance between the 2018 Planck curve and one generated by CLASS using the \texttt{default.ini} settings is approximately 75.  Any curve with a distance to the Planck curve of less than approximately 1000 is visually indistinguishable from it.} Notably, the optimal solution features a negative value for $\Omega_{\Lambda_2}$, a result that finds natural explanation through dimensional regularization in the renormalization procedure.

The methodology described above\footnote{While we use frequentist inference, as noted in \cite{Park:2025iog}, this approach deviates from traditional frequentist methods: cosmological parameters are treated as input variables in the regression framework, with the polynomial coefficients taking on the role of statistical parameters instead of the cosmological parameters themselves. The resulting model serves as a surrogate, approximating the relationship between cosmological parameters and observables.} departs significantly from the standard Bayesian parameter inference employed by the Planck collaboration\cite{Planck:2018vyg}, and several comments are in order to clarify this choice. The primary motivation for adopting a non-standard, deterministic approach lies in the behavior of the newly introduced parameters, particularly 
$\Omega_{\L_2}$. Although theoretical estimates for the magnitude of 
$\Omega_{\L_2}$ are possible and in fact necessary as a preliminary step, it was observed in \cite{Park:2025iog} that these estimates overshoot the best-fit value by several orders of magnitude (a possible explanation for this mismatch is discussed in \cite{Park:2025iog}) - a discrepancy that results in power spectra incompatible with observational data. 
Given the lack of reliable priors and the computational expense of performing full Bayesian inference over such a vast and poorly constrained parameter space, we instead adopted an iterative trial-and-error method. This involved progressively reducing 
$\Omega_{\L_2}$ and performing wide parameter scans until convergence toward the observed Planck 2018 power spectrum was achieved. Although this procedure lacks formal statistical underpinnings (see the conclusion for further discussion), it proved effective in identifying the correct order of magnitude for 
$\Omega_{\L_2}$.

Our analysis leveraged the unbinned theoretical power spectrum file provided by the Planck 2018 collaboration:\\
 \texttt{COM\_PowerSpect\_CMB-base-plikHM-TTTEEE-lowl-lowE-lensing-minimum-theory\_R3.01.txt}.
As this dataset represents a smooth, noise-free theoretical prediction sampled at integer multipoles and lacks an associated covariance matrix, it further justifies our departure from likelihood-based methods that rely on detailed error modeling. In contrast, the observed spectra (given, e.g., in \texttt{COM\_PowerSpect\_CMB-TT-full\_R3.01.txt}) are binned and require covariance information to be properly interpreted, which was not necessary for our purposes.
A major advantage of our approach is its computational efficiency. The entire parameter scan process was executed within the CLASSy framework, with only a minimal external script used to calculate and rank distance measures - an operation that completes nearly instantaneously. In comparison, integrating additional tools like Cobaya or MontePython for full Bayesian inference increased runtime by a factor of six to eight. Given that our data collection alone required over three months, the reduction in computational overhead was a decisive factor. While our trial-and-error method is not a substitute for a rigorous statistical analysis, it enables rapid, resource-efficient exploration of the parameter space. Once a viable range for 
$\Omega_{\L_2}$ is identified, more focused Bayesian scans using Cobaya become both feasible and significantly less time-consuming.

To complement and enhance the efficiency of our parameter space exploration, we implement advanced ANN techniques. Various numerical calculations and machine learning methods have been recently applied to the physics of black hole solutions\cite{Hatefi:2022shc,Hatefi:2023sgr,Hatefi:2023gpj,Hatefi:2024asc}. These machine learning methods enable precise identification of optimal parameter coordinates while maintaining computational tractability across the challenging eight-dimensional parameter space.

\vspace{.3in}

Let us explore the estimation of the distance based on the currently established parameters as well as the finite-T induced ones, $\Omega_{\L_2}$ and $\Omega_{\L_3}$. Throughout this section, the prediction accuracies are translated into the performance of the ANN-based method in detecting the minimum distance to the 2018 Planck curve from statistical populations in the presence of the finite-T effects. 

\begin{figure}[ht]
    \centering
    \includegraphics[scale=0.6]{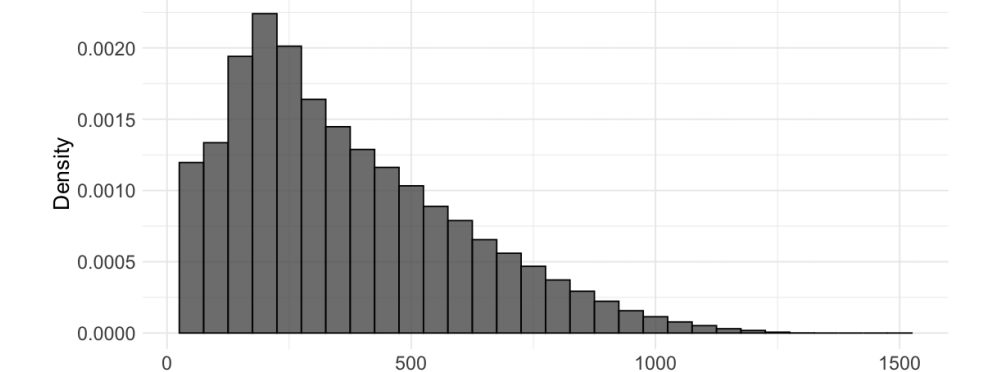}
    \caption{Sampling distribution of the distance values based on 8 cosmological parameters.}
    \label{fig:his_8p}
\end{figure}
The numerical analyses evaluate the distance between the 2018 Planck curve and 7-parameter and 8-parameter ones using 459,912 and 335,985 data points, respectively. These data points are computed over equally spaced grids spanning the domain of the two approaches. These two data sets are considered as two statistical populations corresponding to the two estimation methods. We consider the response variable $y$, representing the distance function $y=f({\bf x})$, which measures the deviation of the proposed estimation methods from the 2018 Planck power spectrum. In this setup, the parameters of the estimation methods were treated as explanatory variables ${\bf x} = (x_1, \ldots, x_J) $, where  $J = 7$ for the 7-parameter model and $J = 8$ for the 8-parameter model. Using this framework, the ML model maps the cosmological parameters to the distance function, enabling us to evaluate and predict the performance of the two estimation methods.
This approach was particularly valuable for assessing the accuracy of the proposed methods, in the presence or absence of $\Omega_{\Lambda_2}$. 

\begin{figure}[ht]
    \centering
    \includegraphics[scale=0.50]{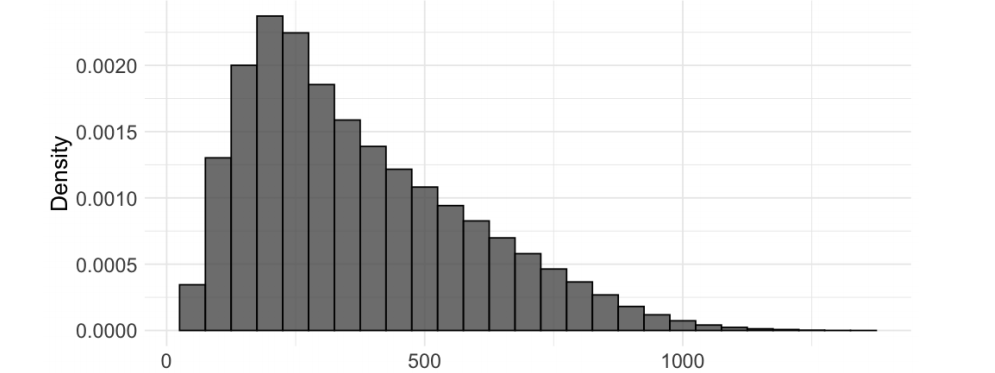}
    \caption{Sampling distribution of the distance values based on 7 cosmological parameters.}
    \label{fig:his_7p}
\end{figure}

Figs. \ref{fig:his_8p} and \ref{fig:his_7p} show the sampling distributions of the distance responses based on two statistical populations derived from the entire dataset in the presence and absence of the leading-order finite-T effect parameter, $\Omega_{\L_2}$, as input data in computation of the distance responses. 
Table \ref{tab:5_numbs} provides descriptive statistics for the distance distributions based on these two statistical populations. While the sampling distributions in Figs. \ref{fig:his_8p} and \ref{fig:his_7p} show that both data capture critical aspects of the distance response function, Table \ref{tab:5_numbs} highlights the clear advantage of the 8-parameter population. It achieves a smaller average distance to the 2018 Planck result, indicating higher accuracy of the 8-parameter estimation method.
Note that this superiority of the 8-parameter method is achieved despite using fewer data points, nearly 100,000 fewer than the 7-parameter competitor and relying on a coarser grid. In contrast, the 7-parameter method, with finer grid sampling and more extensive data, falls short of the prediction accuracy attained by the 8-parameter method. This numerical comparison indicates the effectiveness of our proposed 8-parameter approach, demonstrating its ability to deliver better predictions with less computational effort and fewer resources.

\begin{table}[ht]
    \centering
    \small{
    \begin{tabular}{lcc}
        \toprule
         & Distance population & Distance population \\
        Statistics & with $\Omega_{\L_2}$ and $\Omega_{\L_3}$ & with $\Omega_{\L_3}$ \\
        \midrule
        Size     & 335985 & 459912 \\
        Minimum  & 26.83  & 28.12 \\  
        1st Quartile & 186.97 & 203.91 \\  
        Median  & 313.21 & 321.48 \\  
        Mean & 366.12 & 371.29 \\  
        3rd Quartile  & 510.30 & 505.75 \\  
        Maximum & 1491.66 & 1335.36 \\  
        \bottomrule
    \end{tabular}}
    \caption{The five-number summaries of the distance statistical populations. }
    \label{tab:5_numbs}
\end{table}

\begin{figure}[ht]
    \centering
    \includegraphics[scale=0.6]{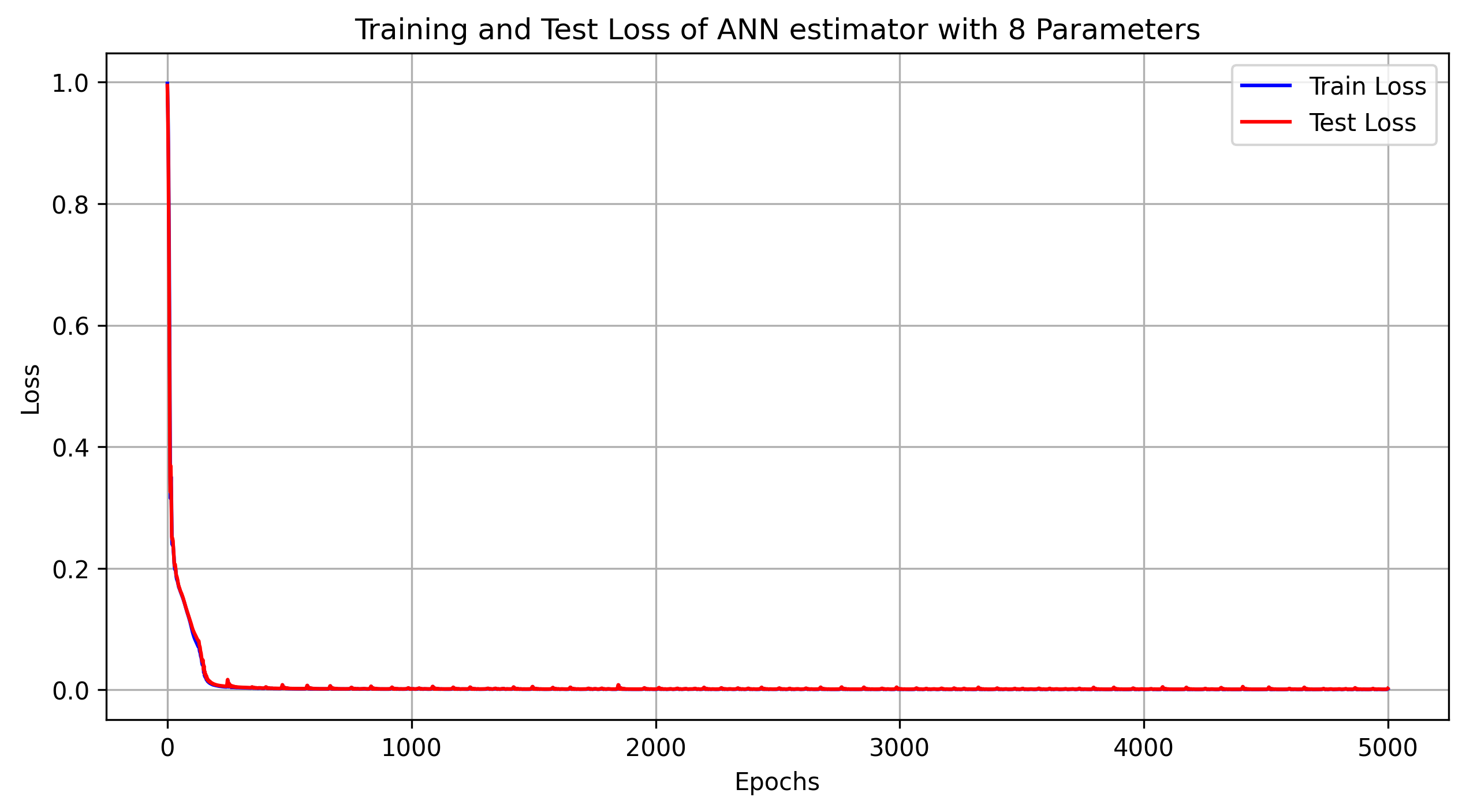}
    \caption{The training and validation loss functions of the proposed ANN method in estimating the distance response function based on eight cosmological parameters with over 5000 epochs.}
    \label{fig:8loss}
\end{figure}
In the following, we treat the output function as the distance between the estimation method and the 2018 Planck data. Hence, the superiority of the methods in predicting the distance function is translated into the superiority of the estimation methods in minimization of the distance response function based on the data generated from the distance function under the two statistical populations. To evaluate the performance of the methods in minimizing the distance response function, we identified 20,000 data points with the smallest distance values from each statistical population. From now on, we will focus on these two data sets as our statistical populations to find the minimum of the distances. Since the functional form of the distance from these statistical populations is unknown, first, one needs to predict the distance as a function of the cosmological features in $\mathbb{R}^7$ and $\mathbb{R}^8$ corresponding to the two statistical populations in absence and presence of $\Omega_{\L_2}$ term, respectively. To do so, we applied ANNs to numerically predict the distance functions from these two statistical populations.

We train the ANNs-based model, described in section \ref{sub:ann} (see Fig. \ref{fig:ann_sd} for a schematic diagram of the trained ANNs), separately for the 7-parameter and 8-parameter statistical populations. 
For the 8-parameter population, we employ an ANN that begins with an input layer of eight nodes, one for each cosmological parameter, followed by four hidden layers, each containing 16 neurons.  We use Rectified Linear Unit (ReLU) activation functions throughout the hidden layers to obtain a robust capability for modelling complex, nonlinear relationships among the parameters in 8D. Before training, all input features are standardized to account for the broad range of magnitudes present in the different coordinates. To do so, each parameter is transformed to zero mean and unit variance. This step facilitates faster convergence and mitigates potential biases introduced by vastly different parameter scales. 
We split the dataset into 80\% training and 20\% test data sets to evaluate prediction performance to validate that model selection and hyperparameter tuning on unseen test data. The network is trained using the Adam optimizer with a learning rate of 0.001 with a mean squared error (MSE) loss function \eqref{ann_loss} over 5000 epochs. For fair comparisons, we train and compute the ANN-based prediction performance of the distance function using a 7-parameter statistics population under the same ANN configuration as described above. By combining a carefully standardized feature set, a structured architecture of hidden layers, and a well-tuned optimization, the ANNs effectively capture the dependencies underlying the distance response in both statistical populations.

\begin{figure}[ht]
    \centering
    \includegraphics[scale=0.6]{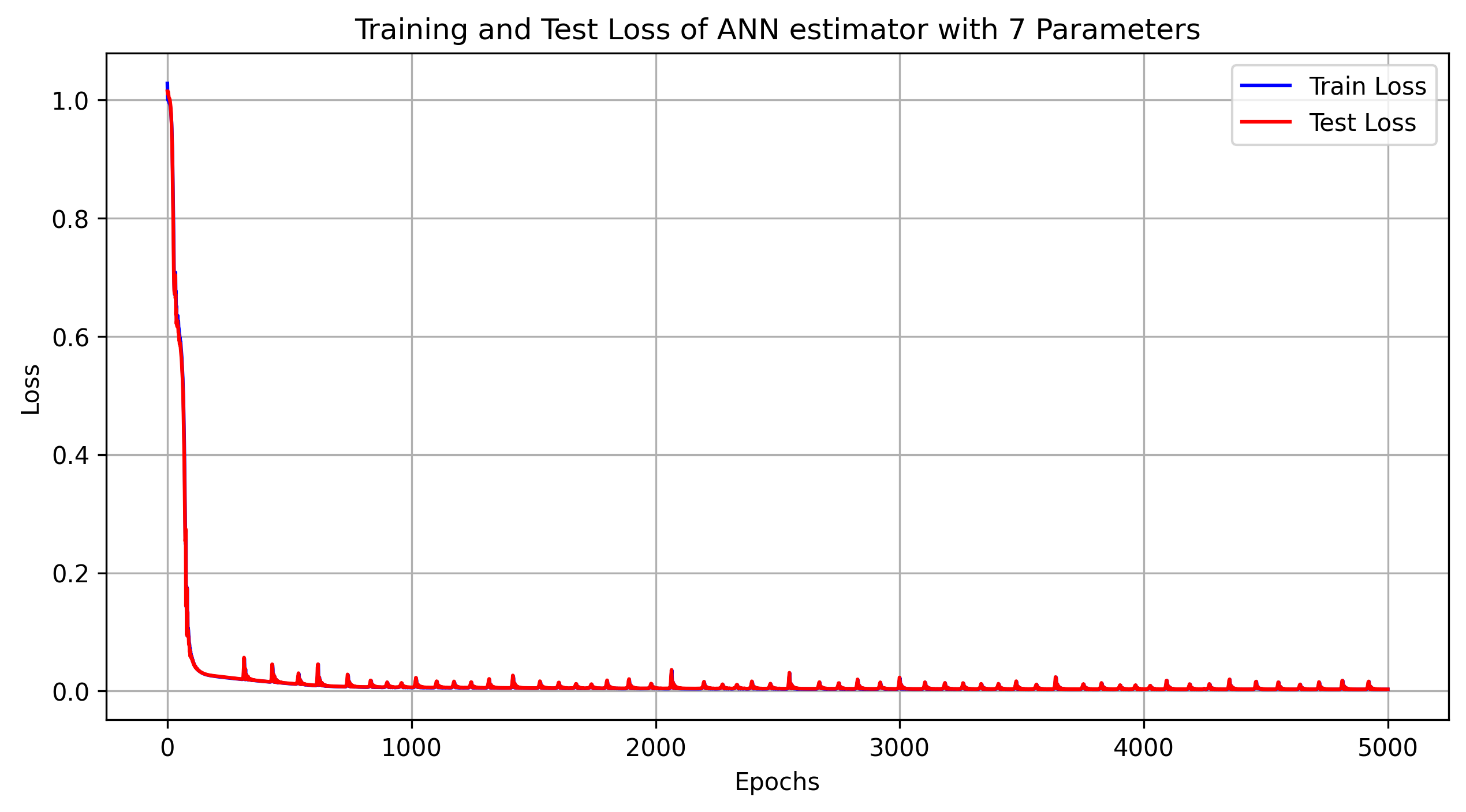}
    \caption{The training and validation loss functions of the proposed ANN method in estimating the distance response function based on seven cosmological parameters with over 5000 epochs.}
    \label{fig:7loss}
\end{figure}

Figs. \ref{fig:8loss} and \ref{fig:7loss} show the training and test loss values based on 8- and  7-parameter statistical populations, respectively. It is easy to see that the training and test loss functions resemble patterns very closely, confirming the ANN-based estimates could predict the underlying distance in both statistical populations. 
Comparison of Figs. \ref{fig:8loss} and \ref{fig:7loss} reveals an interesting contrast. Although the eight-parameter model is constructed from a more coarsely sampled statistical population, its ANN estimator's test losses track the training losses more closely than the seven-parameter model. Consequently, the eight-parameter test loss converges faster, whereas the seven-parameter model shows a slight shortfall, evidenced by larger gaps between its training and test loss curves.

\begin{figure}[ht]
    \centering
    \includegraphics[scale=0.40]{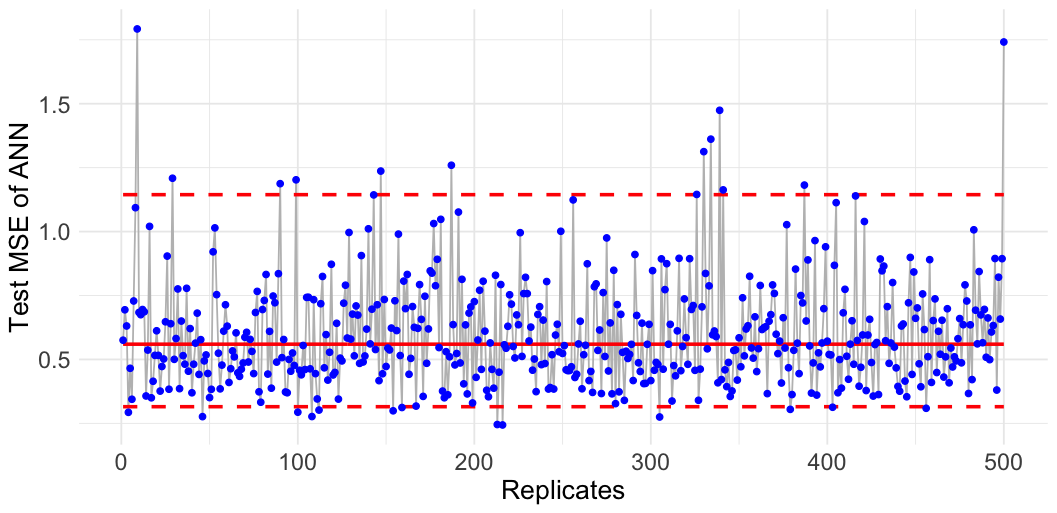}
    \caption{The blue points show the test MSEs of the proposed ANN model for estimating the distance response using eight cosmological parameters based on 500 random replicates. The red dashed lines mark the 95\% confidence interval for the test MSEs (0.31,1.14), and the solid red line shows the median test MSE of 0.55.}
    \label{fig:mse_8p}
\end{figure}

To evaluate the reliability of the proposed ANN model, we conduct 500 independent replications of the ANN fitting procedure, as described earlier, each applying a similar 8-parameter population. In each replication, the model is trained for 1000 epochs, and its predictive performance is assessed by computing the test MSE in estimating the distance function.
The resulting 500 MSE values are then sorted to calculate both the median (50th percentile) and the empirical 95\% confidence interval (CI), determined by the 2.5th and 97.5th percentiles. Fig. \ref{fig:mse_8p} displays all 500 MSEs as blue points, with the median MSE indicated by a solid red line and the 95\% CI shown with dashed red lines. Despite the stochastic nature of ANN, the results demonstrate consistent performance across trials. The test MSEs fluctuate around a median of 0.55, with a relatively narrow 95\% confidence interval ranging from 0.31 to 1.14. These findings highlight the robustness and accuracy of the proposed ANN method in predicting the distance function, even when trained with 1000 epochs.

The gradients to optimize the distance function are computed as described in section \ref{sub:grad}. Subsequently, we employed the BFGS optimization algorithm to find the minimum of the objective function. Since the explicit form of the distance function is unavailable, a trained ANN was used as a surrogate model to approximate the function and compute the gradients numerically. Accordingly, we implemented the gradient-based optimization iteratively until the algorithm converged and the stopping criterion, given by $|| \nabla \mathcal{F}^{L}({\bf x}, \widehat{\boldsymbol{\theta}}_{opt}) ||_2 < 10^{-7}$ is satisfied. (See Appendix B.) Once the ANN-based gradient optimization converges, the estimated optimal coordinates represent the minimum of the approximated distance function. Applying the optimization algorithm to the eight-parameter statistical population, the ANN-based optimization predicted a minimum distance of \( y = 29.24 \) with the corresponding parameter values:
\((\Omega_{\Lambda_2}, \Omega_{\Lambda_3}, h, \omega_b, \omega_{cdm}, A_s, n_s, \tau_{reio}) = (-2.069 \times 10^{-8}, 0.00430, 0.6752, 0.02240, 0.1196, 2.1007 \times 10^{-9}, 0.9660, 0.05424).\)
By using the same optimization configurations but considering the seven-parameter statistical population, the ANN-based gradient optimization leads to a minimum Planck distance of \( y = 32.09 \). The corresponding parameter values at this minimum were:\\
\((\Omega_{\Lambda_3}, h, \omega_b, \omega_{cdm}, A_s, n_s, \tau_{reio}) = (0.0023356, 0.6769, 0.02237, 0.1199, 2.0988 \times 10^{-9}, 0.9658, 0.05396).\)
By comparing the optimization results under the two statistical populations, we observe that the ANN-based estimate of the eight parameter model attains a lower minimum for the distance function. One can then apply the feature ablation method reviewed in section \ref{sub:ablation} to evaluate the relative importance of $\Omega_{\Lambda_2}$, compared to other established cosmological parameters. Since the explicit form of the distance function is not available, we employed feature ablation using ANNs to estimate its impact in two feature importance numerical studies.

We begin by performing an analysis by using ANNs to model the distance function across its entire domain. To construct a comprehensive dataset, we merge the 8-parameter statistical population with a 7-parameter population, setting $\Omega_{\Lambda_2} = 0$ in the latter. This integration resulted in a combined dataset of 795,897 data points.
As described earlier, the ANN architecture consists of four hidden layers with 16 neurons each and is trained for 5000 epochs. The dataset is randomly partitioned into 80\% training and 20\% testing subsets to assess predictive performance. The ANN is then trained to approximate the distance response function over the full dataset. For the ablation study, we remove one input feature at a time, retrain the ANN, and evaluate the resulting change in test MSE. 
Table \ref{tab:mse_metric_feature_removal} presents the test MSEs for both the full model and each reduced model. Despite the high nonlinearity of the distance function ranging from 0 to 1491, the full model achieves an average test MSE of only 4.95, demonstrating the ANN’s strong prediction performance to capture the global structure of the function.
Also, we observe that removing $\Omega_{\Lambda_2}$ led to a greater increase in test MSE compared to removing some of the established parameters, indicating that $\Omega_{\Lambda_2}$ plays a more significant role in predicting the Planck distance across the full domain than those.

\begin{table}[ht]
\centering
\caption{This result is based on the entire 700K data. MSEs of the full models and reduced models after feature removals using ANN models using 5000 epochs based on 700k data points obtained by combining all 8- and 7-parameter statistical populations.}
\vspace{.1in}
\begin{tabular}{cc}
\hline
 Feature Removed &  MSEs using ANNs \\ \hline
 Full Model          & 4.95018   \\
 $\Omega_{\Lambda2}$ & 5.74744   \\
 $\Omega_{\Lambda3}$ & 5.21219    \\
 $h$                 & 189.192   \\
 $\omega_b$          & 595.213   \\
 $\omega_{cdm}$      & 31866.2   \\
 $A_s$               & 35877.0   \\
 $n_s$               & 728.578   \\
 $\tau_{reio}$       & 10252.8   \\ \hline
\end{tabular}
\label{tab:mse_metric_feature_removal}
\end{table}

In the next feature ablation analysis, the aim is to evaluate the individual contributions of each input feature to the minimum values of the distance function. Since the objective is to identify the function's minimum, we focus on a subset of 5000 data points corresponding to the lowest distances from the combined population.
We apply the ANNs by using the same architecture as in the previous analysis, but train with 5000 epochs. For uncertainty quantification, we employ 50-fold cross-validation. In each of the 50 runs, 49 folds (4900 data points) are used for training the ANN, and the remaining fold (100 data points) is used for testing. This process is repeated, so that each fold serves once as the test set. In each run, the test MSE for both the full model and reduced models are calculated, each constructed by removing a single feature. This yields 50 MSE values for the full model and for each reduced model.
\begin{figure}[ht]
    \centering
    \includegraphics[scale=0.54]{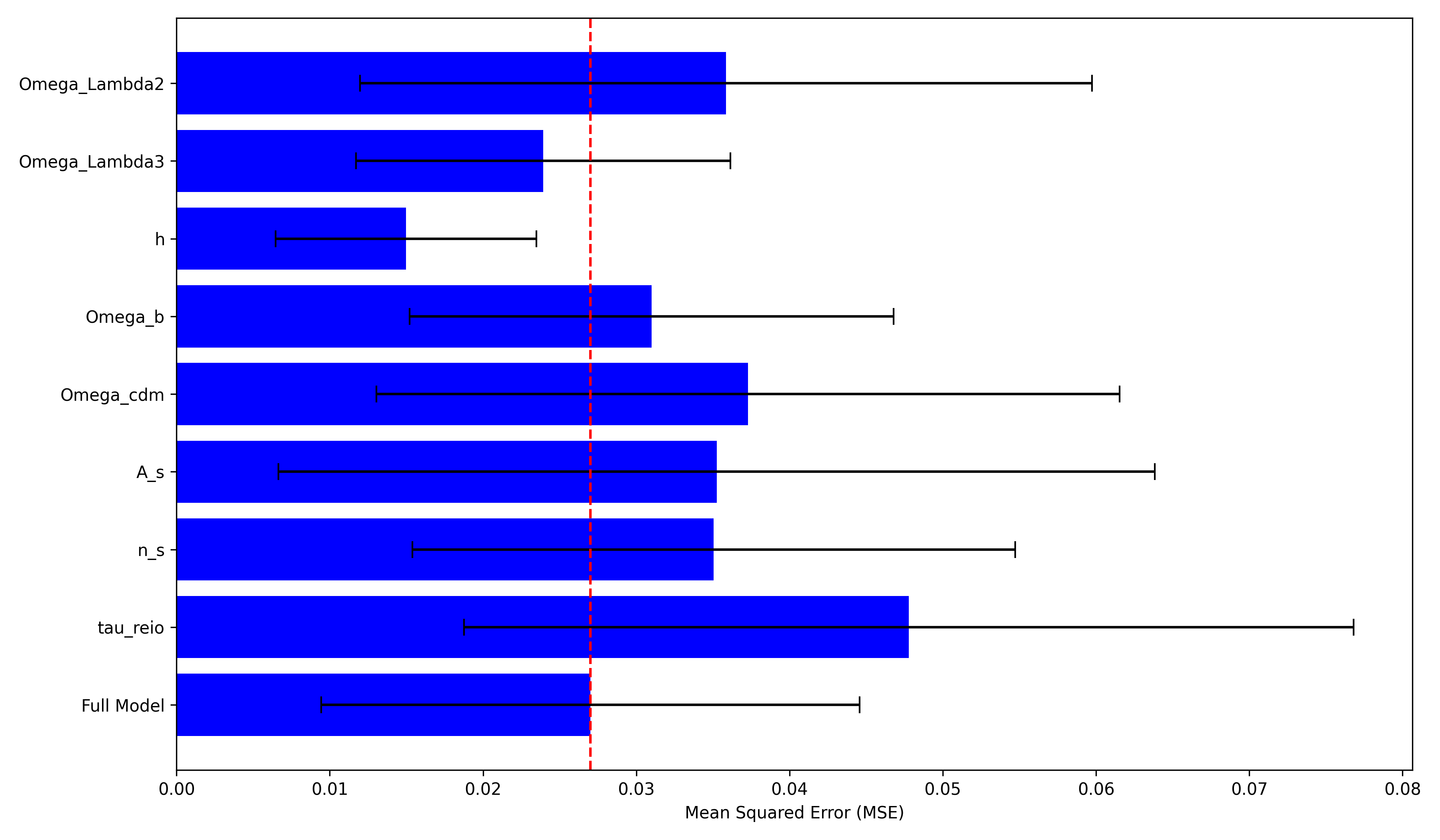}
    \caption{This result is based on the first 5000 shortest distances. Test MSEs of the full and reduced ANN models for predicting the distance function. Each bar represents the mean MSE along with its 95\% confidence interval, computed across 50 folds. Reduced models exclude one input parameter at a time to assess its impact on prediction performance.}
    \label{fig:mse_5k}
\end{figure}
Fig. \ref{fig:mse_5k} presents the results of this analysis, where the height of each blue bar shows the mean test MSE and the black lines denote the 95\% confidence intervals. The variation in MSE across reduced models reveals that the features contribute unequally to the ANN's performance in predicting the minimum distance. In addition, removing the feature $\Omega_{\Lambda_2}$ increased the average test MSE by approximately 25\% compared to the full model, indicating its strong influence. Moreover, when comparing $\Omega_{\Lambda_2}$ to some of the parameters, we observe that the model with $\Omega_{\Lambda_2}$ has a greater impact on model accuracy. 

Figure 7 presents a local sensitivity analysis based on a computationally intensive numerical experiment. Its purpose is to examine the relative importance of each parameter in achieving the minimum distance - essentially identifying which parameters are most critical for finding the 'optimal' values. This analysis focuses specifically on how individual parameters influence the optimization process near that optimum point. Table 3, by contrast, is based on a global analysis using all 700,000 data points across the full parameter space (combining both the 7-parameter and 8-parameter models). The objective of Table 3 is not optimization, but rather to assess which parameters are most important for accurately estimating the entire Planck curve over its full domain in $\mathbb{R}^8$. Due to the large dataset size, the ANN was trained only once for Table 3. Parameter importance was then evaluated using a 'leave-one-out' strategy: each parameter was excluded in turn, and the resulting increase in estimation error was recorded. The magnitude of this error increase serves as a quantitative measure of each parameter’s contribution to the accurate global reconstruction of the Planck curve.

The results shown in Table~\ref{tab:mse_metric_feature_removal} and Fig.~\ref{fig:mse_5k} suggest that $\Omega_{\L_2}$ and $\Omega_{\L_3}$ play an important role in determining the fine structure of the curve. Table~\ref{tab:mse_metric_feature_removal} shows that the large-scale shape of the curve is primarily determined by the standard parameters. However, once attention is restricted to curves that closely match the Planck 2018 result, the parameters $\Omega_{\L_2}$ and $\Omega_{\L_3}$ become essential.

\begin{figure}[ht]
    \centering
    \includegraphics[scale=0.80]{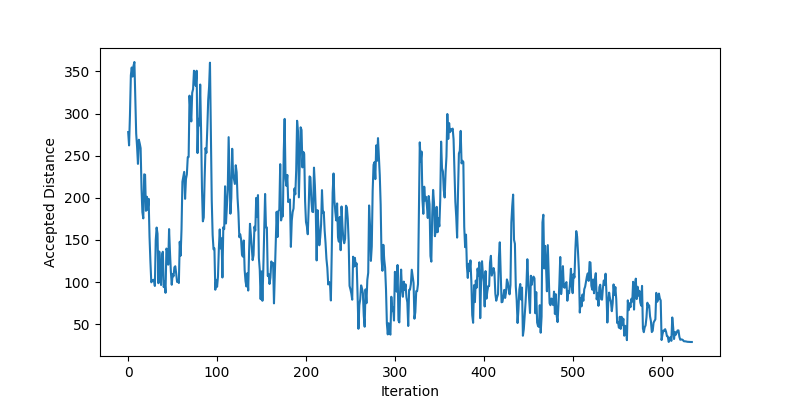}
    \caption{The trace plot of the Markov Chain Monte Carlo of the accepted minimum distances obtained by the stochastic simulated annealing based on ANNs.}
    \label{fig:sim_ann_8p}
\end{figure}

With the significance of $\Omega_{\L_2}$ confirmed through feature ablation, we now use the simulated annealing approach of section \ref{sub:sa}. This method is applied to the statistical population to determine the minimum of the distance function.
The optimization problem involves eight input parameters, with the objective function being unavailable in an explicit analytic form. Instead, it must be approximated stochastically through forward and backward computations of an ANN, as described earlier. Given the integration of multiple numerical methods, it is crucial to account for the sampling variability of parameter estimates and the numerical measurement errors in identifying the minimum. To address this challenge, we employed the stochastic simulated annealing approach within a probabilistic framework. 

The optimization process begins with an initial random solution, and perturbations are introduced iteratively using a random walk Metropolis-Hastings algorithm to explore the parameter space. The initial temperature is set to \( T_0 = 10^4 \), and a cooling schedule is applied with a decay rate of \( \alpha = 0.99 \), gradually reducing the acceptance probability for worse solutions as the search progressed. The ANN predicts the distance response for proposed solutions throughout the optimization, guiding the SA algorithm toward the global minimum. The SA algorithm is run for 20,000 iterations, where 635 transitions are accepted. The results, illustrated in Fig. \ref{fig:sim_ann_8p}, show the MCMC sequence of accepted minimal distances obtained via ANN-based SA. Ultimately, the algorithm converges to the minimum distance value of \( y = 28.8111 \), with the corresponding optimal coordinates:  
\bea
&&(\Omega_{\Lambda_2}, \Omega_{\Lambda_3}, h, \omega_b, \omega_{cdm}, A_s, n_s, \tau_{reio}) \nn\\
&=& (-2.061 \times 10^{-8}, 0.00425, 0.6762, 0.0223, 0.1196, 2.1 \times 10^{-9}, 0.9664, 0.056).
\eea
The  SA algorithm efficiently explores the complex and noisy domain, with the ANN acting as a surrogate model to approximate the distance function. This ensures robustness in parameter estimation, effectively handling both numerical measurement errors in the method and the sampling variability of the estimated parameters.

\section{Conclusion}

In this study, we have incorporated finite-temperature quantum
gravity effects - arising from off-shell vacuum-bubble
contributions of Standard Model fields in the 1PI effective action
- to improve the fit to the CMB power spectrum. These
contributions, which may equivalently be viewed as quantum
corrections to the effective equation of state of the early-Universe
plasma, are parametrized by the density parameters
$\Omega_{\Lambda_2}$ and $\Omega_{\Lambda_3}$ and enter the Hubble
parameter as spatially homogeneous, background-only modifications
scaling as $a^{-4}$ and $a^{-2}$. Our analysis reveals that the
newly introduced density parameter, $\Omega_{\Lambda_2}$, takes on a
negative value - a result that arises naturally within the
framework of renormalization using dimensional regularization.
Meanwhile, the inclusion of the parameter $\Omega_{\Lambda_3}$
introduces additional flexibility in modeling the effective
vacuum-energy sector of the Universe.

We evaluated the effectiveness and robustness of ANN-based methods
in estimating cosmological parameters. Our results demonstrate that
ANNs are capable of capturing complex nonlinear dependencies and
accurately modeling cosmological data. Numerical experiments indicate
that the introduction of finite-temperature-induced density
parameters enhances predictive performance, enabling faster
convergence and improved accuracy, even when training data is
limited. This highlights the potential relevance of quantum
gravitational corrections in refining cosmological predictions.
Notably, our extensive numerical analysis revealed that
finite-temperature corrections can have a more substantial impact on
predictive accuracy than some well-established cosmological
parameters. To further optimize parameter estimation, we incorporated
a simulated annealing algorithm guided by ANN predictions. This
stochastic optimization strategy efficiently navigates the parameter
space while accounting for numerical uncertainties in the distance
function, producing robust estimates for the underlying cosmological
parameters.

We emphasize that the present analysis constitutes a phenomenological
model-building exercise rather than a first-principles derivation.
As discussed in section~\ref{sec:conceptual}, the ideal approach
would be a fully first-principles derivation of the hydrodynamic
system from the underlying finite-temperature quantum field theory
- analogous in spirit to Weinberg's external-field treatment of QED
(Ch.~13.6 of \cite{Weinberg1}) but extended to the full Standard
Model coupled to gravity at finite temperature. Such a derivation
would rigorously establish the separation between on-shell
kinetic-theory content (captured by the standard $\Omega_r$,
$\Omega_{\rm cdm}$, and $\Omega_b$) and off-shell vacuum-bubble
content (captured by $\Omega_{\Lambda_2}$ and $\Omega_{\Lambda_3}$).
The approach adopted here - coupling the 1PI effective action of
the SM+gravity system to a separately specified hydrodynamic sector
- captures the leading effects and serves as a practical starting
point, but is not identical to the first-principles derivation just
described. The clean separation we draw at leading order between
on-shell and off-shell content is the best boundary available at the
current level of approximation, and we have taken care to avoid
double-counting on-shell species. A fully rigorous treatment remains
a genuine open question.

A word is in order regarding the validity of the high-temperature expansion
of $J(m_\phi,T)$ in Eq.~(8), which underlies the $T^4$- and $T^2$-scaling
terms parametrized by $\Omega_{\Lambda_2}$ and $\Omega_{\Lambda_3}$, as
anticipated in the remark following that equation. Strictly speaking, the
high-temperature expansion applies to a species that is light relative to
the ambient temperature, $m \ll T$; for a heavy species, $m \gg T$, one must
instead employ a low-temperature expansion. A fully consistent treatment would
therefore partition the thermal history of the Universe into epochs and
apply, species by species, whichever expansion is valid in each regime.
While correct in principle, this is impractical in the present
phenomenological setting: it would require a major, piecewise restructuring
of the Friedmann sector inside CLASS that would be difficult to implement
reliably.

Two further considerations temper the concern. First, ``high'' and ``low''
temperature are meaningful only relative to the mass of the field, i.e.\ to
the ratio $m/T$. In the present framework $m$ is the \emph{renormalized}
mass, fixed as part of a novel renormalization procedure, so one retains the
freedom to choose the renormalization condition. This is the viewpoint
adopted in our earlier works~\cite{Park:2021ohu,Park:2021vro} on the
cosmological-constant problem, and a related stance appears in \cite{Ageeva:2024qie}. We do not pursue this direction in
detail here; the relevant lesson is that, once such a renormalization
condition is admitted, the boundary between the high- and low-temperature
expansions is not as sharp as it might first appear, and there may well
exist a regime of the renormalized mass for which the high-temperature
expansion remains serviceable and the contributions from the
low-temperature region are correspondingly small. Second, since a
higher-order contribution such as $T^6$ grows rapidly with temperature, its
coefficient must be small in order not to disturb the agreement with the
Planck data already achieved; and because the finite-temperature effect is
naturally strongest in the early, high-temperature Universe, the imprint of
the low-temperature region on integrated observables is expected to be
modest.

We nonetheless acknowledge that the $(T^4,T^2)$ parametrization adopted here,
while natural, is not the whole story. The leading off-shell contribution of
a heavy field, generated by effective interactions obtained upon integrating
it out, scales as $T^6/m^2$ \cite{Matsumoto:1999kd} and would correspond to an additional term
$\sim \Omega_{\Lambda_4}\, a^{-6}$ in the Friedmann equation. A systematic
incorporation of such a term - alongside a full Bayesian inference of the
enlarged parameter set - is a well-motivated direction that we regard as a
priority for follow-up work, resources permitting.

In this regard, we note that higher-order in-medium effects such as Debye screening and collective excitations, which do not follow the naive $T^4$
scaling, represent a natural extension of the present framework and could further refine the effective equation-of-state modelling. Some preliminary groundwork in this direction has been laid in \cite{Park:2024kfn}, where thermal resummation in the gauge and fermion sectors of the Standard Model coupled to gravity is carried out, and the emergence of gauge and fermion thermal masses - directly relevant to Debye screening - is analyzed in detail (sections~3.2 and~3.3 of that work). These subleading contributions, together with a fully systematic treatment of the on-shell/off-shell separation, are natural targets for future investigation.

A central motivation of this work was to explore whether
finite-temperature quantum-gravity corrections - whether framed as
shifts to the cosmological constant or as improved modelling of the
early-Universe plasma equation of state - could help alleviate the
Hubble tension \cite{Riess:2016jrr}. While our findings do not
resolve the tension outright (see, e.g., \cite{SolaPeracaula:2021gxi}
for evidence of resolution in dynamical vacuum models), they suggest
that higher-order finite-temperature effects may play a meaningful
role, warranting further investigation.

A key limitation of our
present statistical analysis is that, although simulated annealing identifies
best-fit parameters, it does not provide likelihood curvature and
therefore cannot yield statistical uncertainties; moreover, because
the distribution of the distance function is unknown, a classical
likelihood cannot be constructed without imposing arbitrary
assumptions.\footnote{In addition, not all digits reported in the
Planck minimum-theory spectrum file
\texttt{COM\_PowerSpect\_CMB-base-plikHM-TTTEEE-lowl-lowE-lensing-%
minimum-theory\_R3.01.txt} are expected to be statistically
significant, which limits the meaningful numerical precision of any
parameter comparison.} To avoid these issues, we reformulated the
problem using an artificial neural network (ANN) trained to
approximate the distance function across the high-dimensional
parameter space in both the 7- and 8-parameter settings, thereby
enabling ANN-based numerical optimization to locate the global
minimum and the associated cosmological parameter estimates without
assuming any particular distribution. To quantify uncertainty and
assess the relative importance of each parameter, we implemented a
numerical ANN-based feature-ablation procedure, which provides 95\%
confidence intervals for the prediction MSE and highlights the
contribution of each cosmological parameter to the distance function;
the resulting intervals are shown in Fig.~7. Looking ahead
\cite{work_in_progress}, implementing a full Bayesian inference
analysis - using packages such as Cobaya or MontePython - will be
a priority for statistically quantifying parameter constraints and
validating the ANN-guided results presented here, and a dedicated
comparison with polarization data will be essential for further
solidifying and refining these conclusions. In particular, such an
analysis would help clarify the independent constraining power of the
CMB on the spatially homogeneous, background-only components
parametrized by $\Omega_{\Lambda_2}$ and $\Omega_{\Lambda_3}$, and
would provide a stringent test of whether these parameters fill a
genuinely unoccupied slot in the standard cosmological parameter
inventory.

\vspace{1in}


\section*{Acknowledgments}

 Armin Hatefi acknowledges the Natural Sciences and Engineering Research Council of Canada  
 (NSERC). Ehsan Hatefi thanks Jose Luis Alvarez-Perez, Pablo Diaz Villar, Philip Siegmann, and Luis Alvarez-Gaume for their helpful discussions and support. Ehsan Hatefi also thanks Enrico Trincherini, Augusto Sagnotti, and the Scuola Normale Superiore in Pisa for their support. Ehsan Hatefi is supported by his International Maria Zambrano Grant from the Ministry of Spain. This project is funded in part by a State of Arkansas Department of Higher Education HIRED Workforce Development Grant, Award No. AWD-105943 \cite{HIRED}. This work used Anvil at Purdue University through allocation
PHY240277 from the Advanced Cyberinfrastructure Coordination
Ecosystem: Services \& Support (ACCESS) program~\cite{ACCESS},
which is supported by U.S. National Science Foundation grants
\#2138259, \#2138286, \#2138307, \#2137603, and \#2138296.

\newpage
\appendix

\renewcommand{\theequation}{A.\arabic{equation}}
\setcounter{equation}{0}

\section{Demonstration of non-degeneracy of $\Omega_{\L_3}$ and $\Omega_K$}

\begin{figure}[ht]
    \centering
    \includegraphics[scale=0.60]{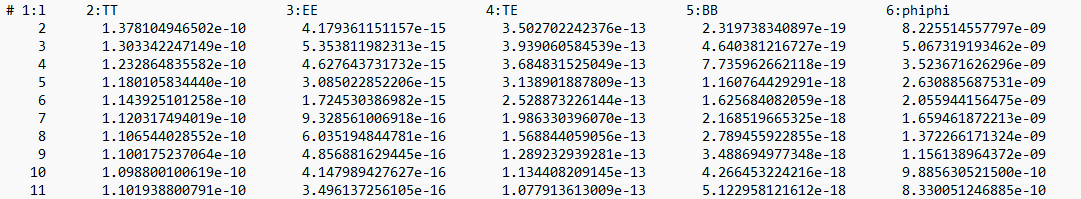}
    \caption{$\Omega_{\L_3}=0.001,\;\Omega_K=0$.}
    \label{OKz}
\end{figure}

\begin{figure}[ht]
    \centering
    \includegraphics[scale=0.60]{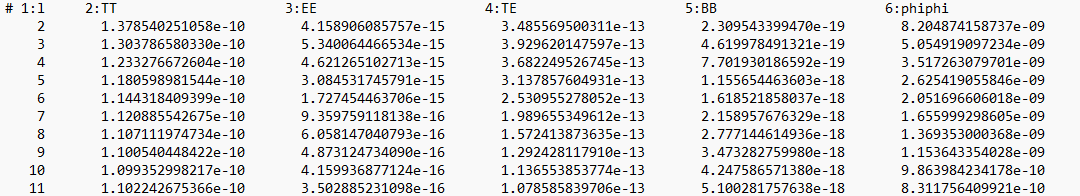}
    \caption{$\Omega_{\L_3}=0,\; \Omega_K=0.001$.}
    \label{OL3z}
\end{figure}

In this appendix, we explicitly demonstrate the non-degeneracy of $\Omega_{\L_3}$ and $\Omega_K$. Let us choose
 $(\Omega_{\Lambda_2}, \Omega_{\Lambda_3}, h, \omega_b, \omega_{cdm}, A_s, n_s, \tau_{reio})$ as follows:
\bea
&& \hspace{2.7in}\Omega_{\L_2}=0.00000017 \nn\\
&&\hspace{-.3in} h\!=\!0.67810,\; \omega_b\!=\!0.02238280,\; \omega_{cdm}\!=\!0.1201075,\; A_s\!=\!2.100549 \!\times \! 10^{-9},\; n_s\!=\!0.9660499,\;  \tau_{reio}\! =\! 0.054308.
\nn\\
\eea
As for $\Omega_{\L_3}, \Omega_K$, let us try the following two combinations:
\bea
&&\Omega_{\L_3}=0.001,\;\Omega_K=0  \nn\\
&&\Omega_{\L_3}=0,\; \Omega_K=0.001.
\eea
Figs. \ref{OKz} and \ref{OL3z} show the first 10 columns of 
the \texttt{cl\_lensed.dat} files obtained by running corresponding \texttt{ini} files.

As these figures show, the entries for the two cases are distinct.

\renewcommand{\theequation}{B.\arabic{equation}}
\setcounter{equation}{0}

\section{Statistical methods}\label{sec:stat}

\begin{figure}[ht]     
\centering
  \includegraphics[scale=0.4]{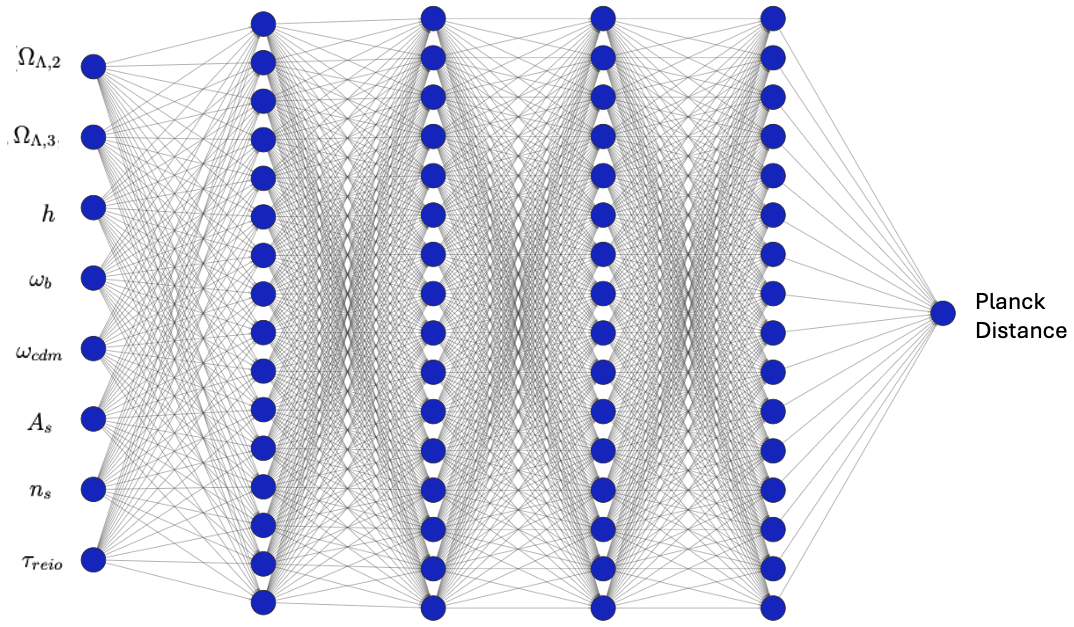}
    \caption{The schematic diagram of the trained ANNs mapping the eight cosmological parameters to the planck distance using four hidden layers, each hidden layer with 16 nodes.}
    \label{fig:ann_sd}
\end{figure}

Let us review the statistical and ML techniques employed in the main analysis. In the present study, we employ ANNs to predict the distance between the 2018 Planck data \cite{Planck:2018vyg} and data generated by the modified CLASS with input cosmological parameter values. This ANN-based model is further used for sensitivity analysis through feature ablation, as well as for probabilistic optimization of cosmological parameters. Fig. \ref{fig:ann_sd} represents the schematic diagram of the ANNs with eight nodes in the input layer representing the eight cosmological parameters and four hidden layers, each with 16 nodes and the output layer with one node of the distance.  Once the ANN is trained, we use gradient-based optimization to determine the set of covariates that minimize the predicted distance.
We then explain the feature ablation strategy to discuss the relative importance of cosmological parameters in the predictions. 
Finally, we developed the ANN-based probabilistic optimization to find the minimum of the distance, considering the numerical measurement errors.

\subsection{Artificial neural networks} \label{sub:ann}

The ANN is a powerful estimation method mapping a set of input explanatory variables (henceforth called features) ${\bf x} = (x_1,\ldots,x_{d_{in}})$ to an output (henceforth called response variable), that is ${\bf y}=(y_1,\ldots,y_{d_{out}})$ \cite{deep_leCun}. The method estimates the relationship between the features and the response variable  ${\bf y}= f({\bf x}), f: \mathbb{R}^{d_{\text{in}}} \rightarrow \mathbb{R}^{d_{\text{out}}}$ through a network of layers consisting of interconnected neurons. 
The network architecture gets information from the input data. The information is then transferred between hidden layers and the activation functions to train the ANN. Once the ANN is trained, the method enables us to learn the relationship between features and response and predict the response function.

Let $\mathcal{F}^{L}(\mathbf{x}, \boldsymbol{\theta})$ represent a deep neural network with $L$ layers that transforms an input from $\mathbb{R}^{d_{\text{in}}}$ into an output in $\mathbb{R}^{d_{\text{out}}}$, where $d_{\text{in}}$ and $d_{\text{out}}$ denote the dimensions of the input and output spaces, respectively.
Suppose $m_l$ represents the number of neurons in the $l$-th hidden layer of the ANN, where $m_0 = d_{\text{in}}$ and $m_L = d_{\text{out}}$ corresponds to the input and output layer sizes, respectively.
Each layer  $l$ $(l=1, \dots, L-1)$ is associated with a weight matrix $\mathbf{W}^l \in \mathbb{R}^{m_l \times m_{l-1}}$, whose element $W^l_{j,k}$ denotes the weight connecting neuron $j$ in layer $l$ to neuron $k$ in layer $l-1$. In addition, each layer uses a bias vector $\mathbf{b}^l \in \mathbb{R}^{m_l}$, where the $j$-th element, $b_j^l$, represents the bias term associated with the $j$-th neuron in layer $l$.
Suppose $\boldsymbol{\theta} = \{ \mathbf{W}^1, \dots, \mathbf{W}^{L}, \mathbf{b}^1, \dots, \mathbf{b}^{L} \}$ denotes the set of all unknown parameters of the ANN.
Let $\mathbf{M}^l$ denote the response from layer $l$ before activation; hence $\mathbf{M}^l$ is given by: 
\begin{equation}
    \mathbf{M}^l = \mathbf{W}^l \cdot \mathcal{F}^{l-1}(\mathbf{x}, \boldsymbol{\theta}) + \mathbf{b}^l, \quad \text{for } l=1, \dots, L-1.
\end{equation}
We use the Rectified Linear Unit (ReLU) activation function  $\sigma(z) = \max(0, z)$ in this work. Applying the activation function $\sigma(\cdot)$, the output of layer $l$ is:
\begin{align}
    \mathcal{F}^{l}(\mathbf{x}, \boldsymbol{\theta}) &= \sigma(\mathbf{M}^l), \quad l=1, \dots, L-1, \\\nonumber
    \mathcal{F}^{L}(\mathbf{x}, \boldsymbol{\theta}) &= \mathbf{M}^L. 
\end{align}
To optimize the ANN's predictive performance, a loss function $\mathcal{J}(\mathcal{F}^{L}(\mathbf{x}, \boldsymbol{\theta}))$ is introduced to quantify the deviation of the predictions from the true target function. Here, the loss function is given by the mean squared loss function  as 
\begin{align}\label{ann_loss}
    \mathcal{J}(\mathcal{F}^{L}(\mathbf{x}, \boldsymbol{\theta})) = \sum_{i=1}^{n} (f(\mathbf{x}_i) - \mathcal{F}^{L}(\mathbf{x},\boldsymbol{\theta}))^2,
\end{align}
where $n$ denotes the size of the training data \cite{elements_hastie}. 
During training, a back-propagation is designed to find optimum solutions $\widehat{\boldsymbol{\theta}}_{opt}$ which minimizes the loss function, that is 
$\widehat{\boldsymbol{\theta}}_{opt} = \arg\min_{\boldsymbol{\theta}} \mathcal{J}(\mathcal{F}^{L}(\mathbf{x}, \boldsymbol{\theta}))$.
The back-propagation algorithm calculates the errors from the output layer backward through the network by using the numerical gradients \cite{Goodfellow:Deep,Rumelhart}. The gradient of the loss with respect to the pre-activation output $\mathbf{M}^l$ in  layer $l$ is computed recursively by
\begin{equation}
    \frac{\partial \mathcal{J}(\mathcal{F}^{L}(\mathbf{x}, \boldsymbol{\theta}))}{\partial \mathbf{M}^l} =
    \left( \mathbf{W}^{l+1} \right)^\top \cdot \frac{\partial \mathcal{J}(\mathcal{F}^{L}(\mathbf{x},  \boldsymbol{\theta}))}{\partial \mathbf{M}^{l+1}} \odot \sigma'(\mathbf{M}^l),
\end{equation}
where $\odot$ denotes the element-wise Hadamard product, and $\sigma'(\cdot)$ is the derivative of the activation function. For the final layer, the gradient is computed by 
\begin{equation}
    \frac{\partial \mathcal{J}(\mathcal{F}^{L}(\mathbf{x},  \boldsymbol{\theta}))}{\partial \mathbf{M}^L} =
    \frac{\partial \mathcal{J}(\mathcal{F}^{L}(\mathbf{x},  \boldsymbol{\theta}))}{\partial \mathcal{F}^{L}(\mathbf{x},  \boldsymbol{\theta})} \odot \sigma'(\mathbf{M}^L).
\end{equation}
One can iteratively update the parameters using stochastic gradient descent to guarantee the convergence of the ANN toward an optimal solution and improve the accuracy of the predictive method.   
Once the ANN algorithm reaches the stationary point, the target response function can be predicted at any new input data ${\bf x}_{\text{new}} $  by 
$\widehat{{\bf y}}_{\text{new}} =  \widehat{f}({\bf x}_{\text{new}}) = \mathcal{F}^{L}({\bf x}_{\text{new}},  \widehat{\boldsymbol{\theta}}_{opt})$
For further details, the readers are referred to \cite{Prince_understand,Goodfellow:Deep}.


\subsection{Gradient-based feature optimization with ANN}\label{sub:grad}

Once the ANN is trained, the next step is to determine the features that minimize the predicted response function ${\bf y}= f({\bf x})$. This process can be reformulated as an optimization problem in which the ANN-based estimate of the objective response function is minimized with respect to the features this time \cite{Nocedal}. 
The gradient of the ANN output with respect to the input features is computed by:
\begin{equation}
    \nabla \mathcal{F}^{L}({\bf x},  \widehat{\boldsymbol{\theta}}_{opt}) = \frac{\partial \mathcal{F}^{L}({\bf x},  \widehat{\boldsymbol{\theta}}_{opt})}{\partial {\bf x}}.
\end{equation}
This gradient is used to iteratively update the features in order to reach the optimal values that minimize the response function. We update the optimum features by: 
\begin{equation}
    {\bf x}^{(t+1)} = {\bf x}^{(t)} - \alpha \nabla \mathcal{F}^{L}({\bf x}^{(t)},  \widehat{\boldsymbol{\theta}}_{opt})
\end{equation}
where $\alpha$ represents the step size tuning parameter used in the gradient descent process.
We employ the Broyden-Fletcher-Goldfarb-Shanno (BFGS) optimization algorithm to find the optimal features \cite{BFGS,Nocedal}. This method updates the input vector iteratively using the following equation:
\begin{equation}
    {\bf x}^{(t+1)} = {\bf x}^{(t)} - {\bf H}_t^{-1}\left( \mathcal{F}^{L}({\bf x}^{(t)},  \widehat{\boldsymbol{\theta}}_{opt})\right) 
     \nabla \mathcal{F}^{L}({\bf x}^{(t)},  
     \widehat{\boldsymbol{\theta}}_{opt}),
\end{equation}
where ${\bf H}_t \left( \mathcal{F}^{L}({\bf x}^{(t)},  \widehat{\boldsymbol{\theta}}_{opt})\right)$ represents an approximation of the Hessian matrix, which ensures a more accurate search direction compared to standard gradient descent. The algorithm is repeatedly alternated until the stopping rule of $|| \nabla \mathcal{F}^{L}({\bf x},  \widehat{\boldsymbol{\theta}}_{opt}) ||_2 < \epsilon$ is satisfied, where $||\cdot||_2$ represents the $\ell_2$ norm.



\subsection{Feature ablation method}\label{sub:ablation}

The feature ablation approach is an established method in ML that determines the relative importance of input features in predictive models. 
The method, as a leave-one-out technique, is applied iteratively to each coordinate in the domain of the objective function to find the contribution of each coordinate in the predictive performance of the underlying ML model \cite{James}. To do that, the feature ablation method iteratively removes each feature and then evaluates the feature's contribution to the performance of the underlying model. 

Let $\{({\bf x}_i,y_i);i=1,\ldots,n\}$ and 
$\{({\bf x}_{j,{\text test}},y_{j,{\text test}});j=1,\ldots,m\}$ denote, respectively, the training and test data of sizes $n$ and $m$ from the model representing the Planck distance response function $y=f({\bf x})$ where $f: \mathbb{R}^{d}  \rightarrow \mathbb{R}$. 
Because the objective response function is not explicitly available, as described in section \ref{sub:ann}, we first predict the objective function using the ANN methods based on the training data. 
Let $\mathcal{F}^{L}({\bf x},  \widehat{\boldsymbol{\theta}}_{opt})$ denote the ANN-based estimate of the objective function $f({\bf x})$. 
The prediction accuracy performance of the trained ANN estimate in the presence of all the input features is computed based on MSE using test data \cite{elements_hastie}. Accordingly, the ANN-based MSE is computed by 
\begin{equation}\label{mse_full_model}
    \text{MSE}\left(\mathcal{F}^{L}({\bf x},  \widehat{\boldsymbol{\theta}}_{opt})\right) = \frac{1}{m} \sum_{j=1}^{m} \left(y_{j,{\text test}} - \mathcal{F}^{L}({\bf x}_{j,{\text test}},  \widehat{\boldsymbol{\theta}}_{opt})\right)^2.
\end{equation}
Henceforth, the MSE \eqref{mse_full_model} is called the full model MSE, showing the prediction error from an ANN-based estimate of the objective function using all the input features. 

In the next step, we iteratively remove the individual feature $x_j, j=1,\ldots,d$
 from the input design matrix. Let ${\bf x}_{-j}$ denote the input features when the  $j$-th feature $x_j$ has been removed from the input data. We then train the ANN estimator of the objective function similar to the full model and find the ANN-based estimate $ \widehat{f({\bf x})} = \mathcal{F}^{L}({\bf x}_{-j},  \widehat{\boldsymbol{\theta}}_{opt})$. Accordingly, we compute the reduced model MSE in a similar vein to equation \eqref{mse_full_model}  using the test data in the absence of the $x_j$ coordinates. Let $\text{MSE}\left(\mathcal{F}^{L}({\bf x}_{-j},  \widehat{\boldsymbol{\theta}}_{opt})\right)$ denote the test MSE of the reduced model using data after removing the $j$-th coordinate for $j=1,\ldots, d$. 
 
We then compute the feature ablation statistic after removing \( x_j \) by
\begin{equation}
    \Delta \text{MSE}_j = 
    \text{MSE}\left(\mathcal{F}^{L}({\bf x}_{-j},  \widehat{\boldsymbol{\theta}}_{opt})\right) -
    \text{MSE}\left(\mathcal{F}^{L}({\bf x},  \widehat{\boldsymbol{\theta}}_{opt})\right), 
\end{equation}
where $\Delta \text{MSE}_j$ statistic shows the contribution of feature $x_j$ in the prediction of the objective function $f({\bf x})$, which can be viewed as the importance of feature $x_j$ in the feature set \cite{elements_hastie}. Therefore, the $\Delta \text{MSE}_j$ statistic can be used as a numerical measure for selection and comparison between features in the prediction of the distance plank response as a function of the cosmological parameters $y=f({\bf x})$. Accordingly, if 
  \( \Delta \text{MSE}_j > 0 \), removing \( x_j \) increases the test MSE. This indicates the relative importance of the $x_j$ feature in the prediction of the objective function.
 When \( \Delta \text{MSE}_j \approx 0 \), removing \( x_j \) has little effect on the performance of the ANN in predicting the objective function.
Finally, when  \( \Delta \text{MSE}_j < 0 \), removing \( x_j \) decreases MSE, suggesting the feature introduces noise in prediction with respect to other features. In other words, the feature $x_j$ is redundant in prediction in the presence of the other variables. 
This feature ablation approach directly measures the significance of input planck parameters in predicting the planck distance function using $\mathcal{F}^{L}({\bf x},  \widehat{\boldsymbol{\theta}}_{opt})$ from the ANN approach.

\subsection{ ANN-based simulated annealing}\label{sub:sa}

 The simulated annealing  (SA) algorithm is an effective numerical method for stochastic optimization problems. 
The goal of the method is to optimize objective function \( y = f({\bf x}) \), where $ {\bf x} \in \mathbb{R}^d $ denotes a set of input covariates. The process follows an iterative procedure where a candidate solution is perturbed at each iteration and then accepted or rejected as the solution through a probabilistic criterion \cite{SA,MC_Casella, Hatefi:2024asc}.

The SA algorithm starts with an initial point \( {\bf x}_0 \) chosen randomly from the sample space and an initial temperature parameter \( T_0 \). The algorithm also requires the cooling rate parameter to calibrate the temperature as the number of iterations increases. 
Like the Markov Chain Monte Carlo (MCMC) statistical method \cite{Andrieu}, the SA algorithm translates the estimation problem (i.e., estimating the optimum point) into two steps. In the first step, the algorithm generates a new candidate for the next iteration of the Markov chain of the solutions. The second step of the algorithm consists of a stochastic step to evaluate if the new candidate should be accepted or rejected as the next solution of the chain. The SA algorithm iteratively assesses whether or not the transition from the current state of the chain to the new candidate improves the optimization of the objective function. 

Here, we explain the implementation of the SA by describing the algorithm's $(k+1)$-th iteration. Let ${\bf x}^{(k)}$ denote the solution of the objective function obtained from the $k$-th iteration of the algorithm. We treat ${\bf x}^{(k)}$ as the current state of the chain and plan to find the next state, that is ${\bf x}^{(k+1)}$.   
Similar to the random walk Metropolis-Hastings method \cite{MC_Casella,Hatefi:2023gpj, Hatefi:2023sgr}, the SA first sample randomly a new candidate ${\bf x}^*$ by random perturbation around the current state of the chain by 
\begin{align*}
    {\bf x}^* = {\bf x}^{(k)} + \Delta {\bf x},
\end{align*} 
where \( \Delta {\bf x} \sim \mathcal{MVN}({\bf 0}, {\bf\Sigma}) \) represents a small random perturbation sampled from a multivariate normal distribution with mean vector ${\bf 0}$ and diagonal variance-covariance matrix  \( {\bf\Sigma} \) whose diagonal entries are the variance parameters for coordinate of the input objective function. These variances can be modified per coordinate to take into account the optimization refinement with respect to each coordinate.

The new candidate ${\bf x}^*$ is then evaluated through a stochastic accept-reject method to see if ${\bf x}^*$ improves the minimum of the objective function or not. Since the objective function $f({\bf x})$ is unknown, the accept reject is applied to the ANN estimate of the objective function $\widehat{f}({\bf x}) = \mathcal{F}^{L}({\bf x},  \widehat{\boldsymbol{\theta}}_{opt})$ computed by the method described in section \ref{sub:ann}. 
Since the closed form of the ANN estimate is not available,  the improvement of the new candidate ${\bf x}^*$  should be evaluated numerically. Hence,  the cost difference based on ANN is computed by 
\begin{equation}
    \Delta \mathcal{F}({\bf x}^*; {\bf x}^{(k)})
    = \mathcal{F}^{L}({\bf x}^*,  \widehat{\boldsymbol{\theta}}_{opt})
     - \mathcal{F}^{L}({\bf x}^{(k)},  \widehat{\boldsymbol{\theta}}_{opt}).
\end{equation}

The next step is to accept or reject the new candidate ${\bf x}^*$.
The probabilistic accept-reject method is designed to accept always ${\bf x}^*$  if it improves the optimization of the objective function \cite{MC_Casella,SA}. However, if ${\bf x}^*$ results in an increase in the cost function, the SA accepts  the new state ${\bf x}^*$ probabilistically through a Metropolis-Hastings (MH) criterion by 
\begin{equation}
    P(\text{accepting the transition from ${\bf x}^{(k)}$ to ${\bf x}^*$}) = \begin{cases} 
        1, & \text{if } \Delta \mathcal{F}({\bf x}^*; {\bf x}^{(k)}) \leq 0, \\
        e^{-\Delta \mathcal{F}({\bf x}^*; {\bf x}^{(k)}) / T_k}, & \text{if } \Delta \mathcal{F}({\bf x}^*; {\bf x}^{(k)}) > 0,
    \end{cases}
\end{equation}
 where \( T_k \) is the current temperature. This probabilistic acceptance allows the algorithm to escape the local minimum and explore the parameter space to find the minimum point more effectively. If the new candidate is accepted, then ${\bf x}^{(k+1)} = {\bf x}^*$; however, if the candidate is rejected, the chain remains in the current state, that is, ${\bf x}^{(k+1)} = {\bf x}^{(k)}$.
 This MH strategy for searching the minimum in the parameter space is refined iteratively by the cooling schedule \cite{Xiang}.
 Here, we apply the exponential decay as the cooling scheme:    
\begin{equation}
    T_{k+1} = \alpha T_k,
\end{equation}
where \( 0 < \alpha < 1 \) is the cooling rate. It is easy to see that in the first iteration of the SA, the high values of the temperature parameter enable the algorithm to explore a wider area of the parameter space to search better for a minimum.   
If the temperature drops too quickly, the algorithm may not be able to search the entire parameter scape and hence may converge to a suboptimal solution. Conversely, if the temperature is cooled too slowly, the algorithm requires many iterations to go from one area to another of the parameter space. Accordingly,  the number of iterations increases, and the $\alpha$ rate controls the cooling temperature for balance exploitation and refinement of the minimum in the parameter space. The algorithm is alternated until no significant improvement is observed in the successive iterations or the algorithm reaches a pre-specified maximum number of iterations.

\end{document}